\documentclass[aps,prb,twocolumn,nofootinbib,showpacs,preprintnumbers,notitlepage,groupedaddress]{revtex4-2}

\usepackage{amsmath,amsfonts,amssymb}
\usepackage[utf8]{inputenc}
\usepackage{braket}
\usepackage{graphicx}
\usepackage{bm}
\usepackage{placeins}
\usepackage{xfrac}
\usepackage{xcolor}
\usepackage[colorlinks=true, breaklinks=true, linkcolor=red, citecolor=blue, urlcolor=blue]{hyperref} 
\usepackage[capitalize]{cleveref}
\usepackage{tikz}
\allowdisplaybreaks

\newcommand{\eff}{\ensuremath{\mathrm{eff}}}
\newcommand{\loc}{\ensuremath{\mathrm{loc}}}
\renewcommand{\S}{\ensuremath{\mathcal{S}}}

\definecolor{tensorblue}{rgb}{0.8,0.8,1}
\definecolor{lighttensorblue}{rgb}{0.9,0.9,1}
\definecolor{tensorred}{rgb}{1,0.5,0.5}
\definecolor{tensorpurp}{rgb}{1,0.5,1}

\tikzset{ten/.style={fill=tensorblue}}
\tikzset{lightten/.style={fill=lighttensorblue}}
\tikzset{tenred/.style={fill=tensorred}}
\tikzset{tengreen/.style={fill=green!50!black!20}}
\tikzset{tenpurp/.style={fill=tensorpurp}}
\tikzset{tengrey/.style={fill=black!20}}
\tikzset{tenorange/.style={fill=orange!30}}
\tikzset{tenwhite/.style={fill=white}}
\tikzset{u/.style={fill=blue!20,draw=black}}
\tikzset{w/.style={fill=green!50!black!80,draw=black}}

\tikzstyle myBG=[line width=3pt,opacity=1.0]

\usepackage{ bbold }

\let\revappendix\appendix

\newcommand{\wh}[1]{\widehat{#1}}

\usetikzlibrary{arrows,positioning} 
\usetikzlibrary{arrows.meta,decorations.markings, snakes}
\usetikzlibrary{math} 
\usepackage{pgf}
\tikzset{
    >=stealth',
    pil/.style={
           ->,
           thick,
           shorten <=2pt,
           shorten >=2pt,}
}

\newcommand\minus{
  \setbox0=\hbox{-}
  \vcenter{
    \hrule width\wd0 height \the\fontdimen8\textfont3
  }
}

\tikzset{->-/.style={decoration={
  markings,
  mark=at position #1 with {\arrow{>}}},postaction={decorate}}}
  
\tikzset{-<-/.style={decoration={
  markings,
  mark=at position #1 with {\arrow{<}}},postaction={decorate}}}

\begin{document}
\title{Direct sampling of projected entangled-pair states}

\author{Tom Vieijra}
\author{Jutho Haegeman}
\author{Frank Verstraete}
\author{Laurens Vanderstraeten}

\affiliation{Department of Physics and Astronomy, Ghent University, B-9000 Ghent, Belgium}

\begin{abstract}
Variational Monte Carlo studies employing projected entangled-pair states (PEPS) have recently shown that they can provide answers on long-standing questions such as the nature of the phases in the two-dimensional $J_1 - J_2$ model.  The sampling in these Monte Carlo algorithms is typically performed with Markov Chain Monte Carlo algorithms employing local update rules, which often suffer from long autocorrelation times and interdependent samples. We propose a sampling algorithm that generates independent samples from a PEPS, bypassing all problems related to finite autocorrelation times. This algorithm is a generalization of an existing direct sampling algorithm for unitary tensor networks. We introduce an auxiliary probability distribution from which independent samples can be drawn, and combine it with importance sampling in order to evaluate expectation values accurately. We benchmark our algorithm on the classical Ising model and on variational optimization of two-dimensional quantum spin models.
\end{abstract}

\maketitle

\section{Introduction}

Strong correlations in quantum many-body systems often lead to interesting phenomena, but are notoriously hard to understand with traditional many-body techniques such as perturbation theory or mean-field approaches. On the computational side, exact diagonalization is necessarily limited to small clusters and quantum Monte-Carlo techniques often suffer from the sign problem. In many cases, the use of variational wavefunctions enables us to get insight into the nature of strongly-correlated ground states. In order to decide on the ground-state properties for a given model Hamiltonian, however, we would like to consider variational classes with a (large) number of parameters that are variationally optimized. In the variational Monte-Carlo (VMC) approach this optimization is performed stochastically as the variational energy is evaluated by a sampling procedure. This VMC approach can be successfully applied for optimizing Jastrow wavefunctions, for describing spin-liquid ground states \cite{Becca2017}, and optimizing variational states inspired by neural networks \cite{Melko2019}.

Tensor networks represent a particularly versatile class of variational states for describing low-energy states of quantum lattice systems. For one-dimensional systems, matrix product states (MPS) are extremely successful, because local expectation values and correlation functions can be computed exactly with low computational cost, and they can be optimized efficiently by the density-matrix renormalization group (DMRG) algorithm or direct variational optimization \cite{Schollwoeck2011}. In two dimensions, projected entangled-pair states (PEPS) \cite{Verstraete2004RenormalizationDimensions} are less efficient: evaluating expectation values comes at a significantly larger computational cost, and finding efficient schemes for the variational optimization is still ongoing \cite{Verstraete2004RenormalizationDimensions, Jordan2008, Jiang2008, Lubasch2014AlgorithmsStates, Phien2015, Vanderstraeten2016, Corboz2016, Liu2017GradientStates, Liao2019, Zaletel2020IsometricDimensions}. Nonetheless, PEPS are, by now, routinely used as a computational tool for simulating ground-state properties of strongly-correlated 2-D quantum systems \cite{Corboz2013, Corboz2014, Zheng2017, Chen2018}, and the framework is being extended to the simulation of thermal states \cite{Czarnik2012, Czarnik2015, Kshetrimayum2017}, excited states \cite{Vanderstraeten2015, Vanderstraeten2019simulating, Ponsioen2020} and time evolution \cite{Czarnik2018, Czarnik2019, Hubig2019}.
\par In that respect, it is natural to consider the use of sampling techniques for evaluating and optimizing PEPS wavefunctions. Although there have been a few works that combine tensor networks with sampling techniques \cite{Sandvik2007VariationalStates, Sandvik2008, Schuch2008SimulationContractions, Wang2010MonteStates, Chou2012, Sikora2015}, only recently stochastic optimization has been put forward as an efficient optimization algorithm for PEPS \cite{Liu2017GradientStates, AccurateLiu2021}; this method showed very strong results for the frustrated $J_1$-$J_2$ Heisenberg model on the square lattice \cite{LiuGaplessModel}. In these works, the Markov-Chain Monte Carlo (MCMC) algorithm was applied to sample PEPS wavefunctions efficiently -- as is standard in the VMC approach for simulating quantum systems.

Recently, however, interest has mounted in sampling procedures that do not involve a Markov Chain, but instead are able to generate a sample directly from the probability distribution. A famous example of these so-called direct sampling approaches is the Box-Muller algorithm to sample variables distributed by Gaussians \cite{Box1958ADeviates}. This approach has the great advantage that all samples can be generated in parallel, which is especially useful in light of recent advances in computer hardware such as massively parallel GPUs.  Furthermore, the samples are independent from each other, which is not the case in MCMC methods.  Related to this interdependency in MCMC are aspects that reduce its efficiency such as long autocorrelation times and poor mixing.  By employing direct sampling schemes, these shortcomings are eliminated. 
Examples of recent progress in direct sampling are flow models \cite{Dinh2016DensityNVP,Kingma2018Glow:Convolutions} and autoregressive neural networks \cite{pmlr-v48-oord16,Oord2016WaveNet} in the context of generative machine learning. These have recently been used in a physical context, such as for sampling classical many-body systems \cite{Wu2019SolvingNetworks}, in the context of lattice field theories \cite{Albergo2019Flow-basedSHANAHAN} and optimization of variational states for many-body systems \cite{Sharir2020DeepSystems, Hibat-Allah2020RecurrentFunctions, casert2020dynamical}.

In the case of MPS, the use of canonical forms can be used to construct a direct sampling algorithm, where the physical degrees of freedom are sampled according to the square of the wave function $|\Psi|^2$ \cite{Ferris2012PerfectNetworks}. The canonical form allows to easily construct low-dimensional conditional distributions, that can be sampled sequentially to generate a configuration from a high-dimensional probability distribution.  Recently, the conditional distributions for classical partition functions modeled as two-dimensional tensor networks were approximated by a tensor renormalization algorithm and the resulting samples were subsequently used as trial states for a Metropolis algorithm \cite{Frias-Perez2021CollectiveRenormalization}.  
Although canonical forms are not generic for PEPS representations and direct sampling according to $p_{\Psi} = |\Psi|^2$ is not possible, in this paper we propose an approximate scheme for generating samples efficiently.  We construct an auxiliary distribution $p_c$, from which we can sample directly, and that is closely related to $p_{\Psi}$.  The mismatch between $p_c$ and $p_{\Psi}$ can be systematically reduced, and is accounted for in the estimation of expectation values by employing importance sampling.  We show that this scheme performs well, and we use this direct sampling approach to efficiently find variational ground states of two-dimensional interacting spin systems.

The paper is organized as follows. In section~\ref{sec:peps}, we review the PEPS construction and how sampling can be used to estimate observables.  In section~\ref{sec:sampling}, we describe the previously used MCMC algorithm, and we introduce our direct sampling algorithm.  In section~\ref{sec:optimization}, we introduce the details of the variational optimization algorithm we use to optimize PEPS wave functions.  In section~\ref{sec:results}, we benchmark our sampling algorithm on the classical Ising model (which can be written as a simple PEPS) and on the variational optimization of the transverse-field Ising model and the antiferromagnetic Heisenberg model.

\section{Projected entangled-pair states}
\label{sec:peps}

We consider quantum spin systems on a two-dimensional square lattice of size $L_x \times L_y$. A classical configuration of spins is denoted as $\S$, and we can associate a quantum state $\ket{\S}$ to each configuration. Any quantum state for this system can be written as a sum over these states with complex weights
\begin{equation}
    \ket{\Psi} = \sum_{\{\mathcal{S}\}} \Psi(\mathcal{S}) \ket{S}.
\end{equation}
In the following, we will also consider subsets of degrees of freedom, for example horizontal rows on the lattice, or single sites, which we will denote as $\mathcal{S}_y$ and $\mathcal{S}_{x,y}$, respectively.  In this paper, we will often sweep through the lattice row by row, going from left to right when sweeping through a specific row.  To denote the partial state we obtained during a sweep up to and including some site with coordinates $(x,y)$, we use the notation $\mathcal{S}_{\leq x, \leq y}$.  The same notation can be used on the level of rows; $\mathcal{S}_{\leq y}$ denotes the state of the first $y$ rows.
\par A PEPS is defined by associating a tensor of order five to each point on the lattice, which can be visualized as
\begin{align}
    A_{\alpha \beta \gamma \delta}^s &= 
    \begin{array}{c}
    \begin{tikzpicture}[scale=.5]
    \draw (-1,0)--(1,0);
    \draw (0,1)--(0,-1);
    \draw (0,0)--(3/4,3/4);
    \node at(-1.2,0) {$\alpha$};\node at(0,-1.4) {$\delta$};\node at(0,1.4) {$\beta$};
    \node at(1.2,0) {$\gamma$};
    \filldraw[ten] (-1/2,-1/2)--(1/2,-1/2)--(1/2,1/2)--(-1/2,1/2)--(-1/2,-1/2);
    \node at(1,1) {$s$};
    \end{tikzpicture}
    \end{array},
    \label{eq:tensor}
\end{align}
where the indices $\alpha$, $\beta$, $\gamma$ and $\delta$ are the virtual indices and the index $s$ denotes a physical degree of freedom, for example the up and down states of a spin degree of freedom. The PEPS tensor network is defined by contracting the virtual indices with those of the tensors associated to nearest neighbors on the lattice, visualized as 
\begin{equation}
\ket{\Psi(\{A\})} = 
\begin{array}{c}
\begin{tikzpicture}[scale=.5]
\foreach \x in {0,2,...,6}{
\draw[shift={(\x,0)}] (0,-1)--(0,7);
\draw[shift={(0,\x)}] (-1,0)--(7,0);
};
\foreach \x in {0,2,...,6}{
    \foreach \y in {0,2,...,6}{
    	\draw[shift={(\x,\y)}]  (0,0)--(3/4,3/4);
    	\filldraw[ten,shift={(\x,\y)}]  (-1/2,-1/2)--(1/2,-1/2)--(1/2,1/2)--(-1/2,1/2)--(-1/2,-1/2);
	};
};
\end{tikzpicture}
\end{array},
\label{eq:pepsstate}
\end{equation}
where the connected lines imply a tensor contraction. In Eq.~\eqref{eq:pepsstate}, the open virtual lines at the boundary are chosen to have dimension 1, and will be dropped hereafter. 
\par In general, computing the norm or an expectation value for a PEPS cannot be done exactly with limited computational resources \cite{Schuch2007}, and approximation techniques have to be considered. Representing the norm of a PEPS as
\begin{equation}
\braket{\Psi(\{A\})|\Psi(\{A\})} =
\begin{array}{c}
\begin{tikzpicture}[scale=.5]
\foreach \x in {0,2,...,6}{
\draw[shift={(\x+3/4,0+3/4)}] (0,0)--(0,6);
\draw[shift={(0+3/4,\x+3/4)}] (0,0)--(6,0);
\draw[color=white,myBG,shift={(\x,0)}] (0,0) -- (0,6);
\draw[shift={(\x,0)}] (0,0) -- (0,6);
\draw[color=white,myBG,shift={(0,\x)}] (0,0) -- (6,0);
\draw[shift={(0,\x)}] (0,0)--(6,0);
};
\foreach \x in {0,2,...,6}{
    \foreach \y in {0,2,...,6}{
    	\filldraw[lightten,shift={(\x+3/4,\y+3/4)}]  (-1/2,-1/2)--(1/2,-1/2)--(1/2,1/2)--(-1/2,1/2)--(-1/2,-1/2);    	
    	\draw[shift={(\x,\y)}]  (0,0)--(3/4,3/4);
    	\filldraw[ten,shift={(\x,\y)}]  (-1/2,-1/2)--(1/2,-1/2)--(1/2,1/2)--(-1/2,1/2)--(-1/2,-1/2);
	};
};
\end{tikzpicture}
\end{array},
\label{eq:double_layer}
\end{equation}
we can imagine that we can move through the lattice from the top to the bottom and progressively multiply the row-to-row transfer matrix. Since the dimension of the boundary operator scales exponentially with the number of columns, we can approximate it as a matrix product operator -- a boundary MPO -- with a tractable bond dimension $\chi_d$. In each step of this row-to-row contraction \cite{Verstraete2004RenormalizationDimensions},
\begin{multline}
\begin{array}{c}
\begin{tikzpicture}[scale=.5]
\draw[shift={(0+3/4,0+3/4)}] (0,0)--(6,0);
\draw[shift={(0+3/8,2)}] (0,0)--(6,0);
\foreach \x in {0,2,4,6}{
\draw[shift={(\x+3/4,0+3/4)}] (0,-1)--(0,0);
\draw[shift={(\x,0)}] (0,-1) -- (0,0);
};
\draw[color=white,myBG,shift={(0,0)}] (0,0) -- (6,0);
\draw[shift={(0,0)}] (0,0)--(6,0);
\foreach \x in {0,2,4,6}{
    \draw (\x+3/8,2) to[out=0, in=90, distance=0.25cm] (\x+3/4,1);
    \draw[color=white,myBG,shift={(0,0)}] (\x+3/8,2) to[out=0, in=90, distance=-0.5cm] (\x,0);
    \filldraw[tengreen,shift={(\x+3/8,2)}]  (-1/2,-1/2)--(1/2,-1/2)--(1/2,1/2)--(-1/2,1/2)--(-1/2,-1/2);    
    \draw (\x+3/8,2) to[out=0, in=-90, distance=-0.5cm] (\x,0);
    \foreach \y in {0}{
    	\filldraw[lightten,shift={(\x+3/4,\y+3/4)}]  (-1/2,-1/2)--(1/2,-1/2)--(1/2,1/2)--(-1/2,1/2)--(-1/2,-1/2);    	\draw[shift={(\x,\y)}]  (0,0)--(3/4,3/4);
    	\filldraw[ten,shift={(\x,\y)}]  (-1/2,-1/2)--(1/2,-1/2)--(1/2,1/2)--(-1/2,1/2)--(-1/2,-1/2);
	};
};
\end{tikzpicture}
\end{array} \\
=
\begin{array}{c}
\begin{tikzpicture}[scale=.5]
\draw[shift={(0+3/8,2)},line width=2] (0,0)--(6,0);
\foreach \x in {0,2,4,6}{
    \draw (\x+3/8,2) to[out=0, in=90, distance=0.25cm] (\x+3/4,1+1/4);
    \filldraw[tengreen,shift={(\x+3/8,2)}]  (-1/2,-1/2)--(1/2,-1/2)--(1/2,1/2)--(-1/2,1/2)--(-1/2,-1/2);
    \draw (\x+3/8,2) to[out=0, in=-90, distance=-0.5cm] (\x,1-1/4);
}
\end{tikzpicture}
\end{array},
\label{eq:double_layer_transfer}
\end{multline}
the bond dimension of the boundary MPO grows. Therefore, in order to keep the bond dimension $\chi_d$ tractable, we approximate the boundary MPO by an MPO with a tractable bond dimension
\begin{multline}
\begin{array}{c}
\begin{tikzpicture}[scale=.5]
\draw[shift={(0+3/8,2)},line width=2] (0,0)--(6,0);
\foreach \x in {0,2,4,6}{
    \draw (\x+3/8,2) to[out=0, in=90, distance=0.25cm] (\x+3/4,1+1/4);
    \filldraw[tengreen,shift={(\x+3/8,2)}]  (-1/2,-1/2)--(1/2,-1/2)--(1/2,1/2)--(-1/2,1/2)--(-1/2,-1/2);
    \draw (\x+3/8,2) to[out=0, in=-90, distance=-0.5cm] (\x,1-1/4);
}
\end{tikzpicture}
\end{array} 
\\ \approx
\begin{array}{c}
\begin{tikzpicture}[scale=.5]
\draw[shift={(0+3/8,2)}] (0,0)--(6,0);
\foreach \x in {0,2,4,6}{
    \draw (\x+3/8,2) to[out=0, in=90, distance=0.25cm] (\x+3/4,1+1/4);
    \filldraw[tengreen,shift={(\x+3/8,2)}]  (-1/2,-1/2)--(1/2,-1/2)--(1/2,1/2)--(-1/2,1/2)--(-1/2,-1/2);
    \draw (\x+3/8,2) to[out=0, in=-90, distance=-0.5cm] (\x,1-1/4);
}
\end{tikzpicture}
\end{array}.
\end{multline}
This approximation step can be performed by local truncation steps or by solving a variational optimization problem, the complexity of which scales typically as $\mathcal{O}(\chi_d^2D^6+\chi_d^3D^4)$.
\par In the spirit of the DMRG algorithm for MPS, the variational optimization of PEPS wavefunctions can be performed by a sweeping algorithm \cite{Verstraete2004RenormalizationDimensions}, where the PEPS tensors are updated consecutively until convergence. In the absence of a general prescription for bringing a PEPS into canonical form, the local optimization of a single tensor, however, suffers from an ill-conditioned inversion problem and is, therefore, unstable. Performing imaginary-time evolution with a Trotter-Suzuki splitting of the time-evolution operator \cite{Lubasch2014UnifyingContractions, Lubasch2014AlgorithmsStates} suffers from the same poor conditioning. Global optimization strategies are potentially more stable, but computing the gradient for the energy function is not straightforward. For infinite PEPS, gradient-based optimization strategies \cite{Vanderstraeten2016} have indeed shown to be more stable and yield states that are variationally optimal \cite{Corboz2016, Vanderstraeten2016}.

\section{Sampling PEPS}
\label{sec:sampling}

Instead of considering the double layer and contracting it using a boundary MPO, another approach consists of interpreting the contraction over the physical indices of the PEPS in Eq.~\eqref{eq:double_layer} as an explicit sum over the set of classical configurations
\begin{align}
\braket{\Psi(\{A\})|\Psi(\{A\})} &= \sum_{\{\mathcal{S}\}} |\braket{\S|\Psi(\{A\})}|^2 \nonumber  \\
&=
\begin{array}{c}
\begin{tikzpicture}[scale=.5]
\foreach \x in {0,2,...,6}{
\draw[shift={(\x+3/4,0+3/4)}] (0,0)--(0,6);
\draw[shift={(0+3/4,\x+3/4)}] (0,0)--(6,0);
\draw[color=white,myBG,shift={(\x,0)}] (0,0) -- (0,6);
\draw[shift={(\x,0)}] (0,0) -- (0,6);
\draw[color=white,myBG,shift={(0,\x)}] (0,0) -- (6,0);
\draw[shift={(0,\x)}] (0,0)--(6,0);
};
\foreach \x in {0,2,...,6}{
    \foreach \y in {0,2,...,6}{
    	\filldraw[lightten,shift={(\x+3/4,\y+3/4)}]  (-1/2,-1/2)--(1/2,-1/2)--(1/2,1/2)--(-1/2,1/2)--(-1/2,-1/2);    	\draw[shift={(\x,\y)}]  (0,0)--(3/4,3/4);
    	\filldraw[tenred,shift={(\x+2/4,\y+2/4)}] (0,0) circle (0.150);
    	\filldraw[ten,shift={(\x,\y)}]  (-1/2,-1/2)--(1/2,-1/2)--(1/2,1/2)--(-1/2,1/2)--(-1/2,-1/2);
	};
};
\end{tikzpicture}
\end{array},
\label{eq:sum_double_layer}
\end{align}
where we define
\begin{align}
\mathbb{1} &=
\begin{array}{c}
\begin{tikzpicture}[scale=.5]
    	\draw (-2/4,-2/4)--(2/4,2/4);
    	\filldraw[tenred] (0,0) circle (0.150);
\end{tikzpicture}
\end{array} = \sum_{\{s\}}
\begin{array}{c}
\begin{tikzpicture}[scale=.5]
    	\draw (-3/4,-3/4)--(-1/4,-1/4);
    	\draw (1/4,1/4)--(3/4,3/4);
    	\filldraw[tenwhite] (1/4,1/4) circle (0.40);
    	\filldraw[tenwhite] (-1/4,-1/4) circle (0.40);
    	\node at(1/4,1/4) {$s$};
    	\node at(-1/4,-1/4) {$s$};
\end{tikzpicture}
\end{array}.
\end{align}
The number of terms in the sum in Eq.~\eqref{eq:sum_double_layer} scales exponentially as a function of the number of physical degrees of freedom. However, if the state is normalized, the terms in this sum are probabilities corresponding to the square amplitudes of the wave function and the sum in Eq.~\eqref{eq:sum_double_layer} corresponds to the normalization of this probability distribution.  Furthermore, when considering expectation values of (sparse) operators they can be rewritten as a classical expectation value over the probability distribution $|\Psi(\mathcal{S})|^2$
\begin{align}
    \langle \widehat{O} \rangle &= \frac{\braket{\Psi|\widehat{O}|\Psi}}{\braket{\Psi|\Psi}} \nonumber \\
    &= \sum_{\{\mathcal{S}\}} \frac{|\Psi(\mathcal{S})|^2}{\braket{\Psi|\Psi}} \frac{\braket{\mathcal{S} | \widehat{O} | \Psi}}{\Psi(\mathcal{S})} \nonumber \\
    &= \sum_{\{\mathcal{S}\}} \frac{|\Psi(\mathcal{S})|^2}{\braket{\Psi|\Psi}} O_{\loc}(\mathcal{S}), 
\label{eq:expectation}
\end{align}
where we introduce the so-called local estimator of the operator $\widehat{O}$
\begin{equation}
O_{\loc}(\mathcal{S}) = \braket{\mathcal{S} | \widehat{O} | \Psi}/\Psi(\mathcal{S}),
\end{equation}
and 
\begin{equation}
    p_\Psi(\S) \equiv \frac{|\Psi(\S)|^2}{\braket{\Psi|\Psi}}
\end{equation}
can be interpreted as a probability distribution. Using sampling algorithms, this expectation value can be estimated by sampling this distribution and evaluating the mean of $O_{\loc}(\mathcal{S})$ for all drawn samples
\begin{equation}
    \langle \widehat{O} \rangle \approx \frac{1}{N} \sum_{\mathcal{S} \sim p_{\Psi}} O_{\loc}(\mathcal{S}).
\label{eq:expectation_sampled}
\end{equation}
The variance on the expectation value in Eq.~\eqref{eq:expectation_sampled} is then
\begin{equation}
    \textrm{Var}_{p_{\Psi}}(\langle \widehat{O} \rangle) = \left< \left( \frac{1}{N} \sum_{\mathcal{S} \sim p_{\Psi}} ( O_{\loc}(\mathcal{S}) - \widehat{O}) \right)^2 \right>.
\label{eq:var_psi}
\end{equation}
Note that in Eqs.~\eqref{eq:expectation} and \eqref{eq:expectation_sampled}, only amplitudes of the wave function appear
\begin{align}
\Psi(\mathcal{S}) = 
\begin{array}{c}
\begin{tikzpicture}[scale=.5]
\foreach \x in {0,2,...,6}{
\draw[shift={(\x,0)}] (0,0) -- (0,6);
\draw[shift={(0,\x)}] (0,0)--(6,0);
};
\foreach \x in {0,2,...,6}{
    \foreach \y in {0,2,...,6}{
        \draw (\x+1/4,\y+1/4)--(\x+4/4,\y+4/4);
    	\filldraw[tenwhite] (\x+4/4,\y+4/4) circle (0.40);
    	\node at(\x+4/4,\y+4/4) {$s$};
    	\filldraw[ten,shift={(\x,\y)}]  (-1/2,-1/2)--(1/2,-1/2)--(1/2,1/2)--(-1/2,1/2)--(-1/2,-1/2);
	};
};
\end{tikzpicture}
\end{array}.
\label{eq:single_layer}
\end{align}
In contrast to the double-layer approach of Eq.~\eqref{eq:double_layer}, here we need to contract a \emph{single-layer} PEPS. Similar to the double-layer contraction of Eq.~\eqref{eq:double_layer_transfer}, we can traverse the lattice row by row and multiply the transfer matrix with a boundary vector -- in this case a boundary MPS -- which is kept at a tractable bond dimension $\chi_s$:
\begin{align}
& \begin{array}{c}
\begin{tikzpicture}[scale=.5]
\draw[shift={(0,2)}] (0,0)--(6,0);
\foreach \x in {0,2,4,6}{
\draw[shift={(\x,0)}] (0,-1) -- (0,0);
};
\draw[shift={(0,0)}] (0,0)--(6,0);
\foreach \x in {0,2,4,6}{
    \draw (\x,2) -- (\x,0);
    \filldraw[tengreen,shift={(\x,2)}]  (-1/2,-1/2)--(1/2,-1/2)--(1/2,1/2)--(-1/2,1/2)--(-1/2,-1/2);    
    \foreach \y in {0}{
    	\draw (\x+1/4,\y+1/4)--(\x+4/4,\y+4/4);
    	\filldraw[tenwhite] (\x+4/4,\y+4/4) circle (0.40);
    	\node at(\x+4/4,\y+4/4) {$s$};
    	\filldraw[ten,shift={(\x,\y)}]  (-1/2,-1/2)--(1/2,-1/2)--(1/2,1/2)--(-1/2,1/2)--(-1/2,-1/2);
	};
};
\end{tikzpicture}
\end{array} \nonumber \\
= & \qquad\qquad \begin{array}{c}
\begin{tikzpicture}[scale=.5]
\draw[shift={(0,2)},line width=2] (0,0)--(6,0);
\foreach \x in {0,2,4,6}{
    \draw (\x,2) -- (\x,1);
    \filldraw[tengreen,shift={(\x,2)}]  (-1/2,-1/2)--(1/2,-1/2)--(1/2,1/2)--(-1/2,1/2)--(-1/2,-1/2);
}
\end{tikzpicture}
\end{array} \nonumber \\
\approx 
&\qquad\qquad \begin{array}{c}
\begin{tikzpicture}[scale=.5]
\draw[shift={(0,2)}] (0,0)--(6,0);
\foreach \x in {0,2,4,6}{
    \draw (\x,2) -- (\x,1);
    \filldraw[tengreen,shift={(\x,2)}]  (-1/2,-1/2)--(1/2,-1/2)--(1/2,1/2)--(-1/2,1/2)--(-1/2,-1/2);
}
\end{tikzpicture}
\end{array}.
\label{eq:single_layer_transfer}
\end{align}
Contracting only single-layer PEPSs can be performed with a significant decrease in computational cost as compared to the double-layer contraction; in particular, it only scales as $\mathcal{O}(\chi_s^2D^4 + \chi_s^3D^2)$.  Furthermore, it has been pointed out that $\chi_s$ can be chosen a lot smaller than $\chi_d$~\cite{Liu2017GradientStates}.  However, finding a relation between the two, if at all present, is still an open question as these bond dimensions are strongly dependent on the model and the application. Yet, despite the scaling advantage for contracting a single layer PEPS, it has to be noted that this single layer contraction needs to be performed for many different samples $\mathcal{S}$, as can be seen in Eq.~\eqref{eq:expectation_sampled}. This is in contrast to the calculation of expectation values in a double-layer PEPS algorithm, where the expectation value can be expressed as a single contraction.

In the remainder of this section, we explain how to estimate observables such as in Eq.~\eqref{eq:expectation} using sampling.  First, we review the Markov Chain Monte Carlo method, which yields samples $\mathcal{S}$ distributed according to $p_{\Psi}(\mathcal{S})$. Afterwards we introduce an algorithm that yields samples $\mathcal{S}$ distributed according to a distribution $p_{c}(\mathcal{S})$, which lies close to the distribution $p_{\Psi}(\mathcal{S})$.  Using importance sampling with the distribution $p_{c}(\mathcal{S})$, the observable in Eq.~\eqref{eq:expectation} can be estimated.  Converse to MCMC, sampling from $p_{c}(\mathcal{S})$ can be performed directly and we introduce a way to systematically bring $p_{c}(\mathcal{S})$ closer to $p_{\Psi}(\mathcal{S})$, when desired.

\subsection{Markov-chain Monte Carlo}

As is standard in the VMC approach, a straightforward way of sampling PEPS tensor networks is using a Markov-chain Monte Carlo algorithm (MCMC). An example is the Metropolis-Hastings algorithm~\cite{Metropolis1953EquationMachines,Hastings1970MonteApplications}, which provides a sampling procedure that requires few assumptions of the underlying probability distribution. In the Metropolis-Hastings algorithm, starting from an initial configuration $\mathcal{S}^1$, one proposes a new configuration $\mathcal{S}^2$ by perturbing $\mathcal{S}^1$ according to a trial probability $\omega(\mathcal{S}^1 \rightarrow \mathcal{S}^2)$.  The configuration $\mathcal{S}^2$ is accepted with probability 
\begin{multline}
P_{\mathrm{acc}}(\mathcal{S}^1 \rightarrow \mathcal{S}^2) \\= \textrm{min}\left(1,\frac{\left|\Psi(\mathcal{S}^2)\right|^2 \omega(\mathcal{S}^2 \rightarrow \mathcal{S}^1)}{\left|\Psi(\mathcal{S}^1)\right|^2\omega(\mathcal{S}^1 \rightarrow \mathcal{S}^2)} \right).
\label{eq:acceptance}
\end{multline}
The conditional probability to transition from state $\mathcal{S}^1$ to $\mathcal{S}^2$ is then given by
\begin{equation}
    T(\mathcal{S}^1 \rightarrow \mathcal{S}^2) = \omega(\mathcal{S}^1 \rightarrow \mathcal{S}^2) P_{\mathrm{acc}}(\mathcal{S}^1 \rightarrow \mathcal{S}^2).
    \label{eq:transition}
\end{equation}
Iterating this procedure yields a chain of configurations that produces samples from the distribution $\left|\Psi(\mathcal{S})\right|^2$ in the limit of infinite number of samples.  This is guaranteed by the detailed balance condition for the trial probabilities
\begin{equation}
\left|\Psi(\mathcal{S}^1)\right|^2 T(\mathcal{S}^1 \rightarrow \mathcal{S}^2) = \left|\Psi(\mathcal{S}^2)\right|^2 T(\mathcal{S}^2 \rightarrow \mathcal{S}^1),
\label{eq:detailed_balance}
\end{equation}
which is in this case implied by the acceptance probabilities in Eq.~\eqref{eq:transition}.
A clear disadvantage is that the transition between two configurations implies a degree of correlation between subsequent configurations.
The choice of $\omega(\mathcal{S}^1 \rightarrow \mathcal{S}^2)$ determines the autocorrelation time and mixing dynamics, and hence the efficiency of the sampling algorithm.  In VMC algorithms, the choice for $\omega(\mathcal{S}^1 \rightarrow \mathcal{S}^2)$ is typically a local update, involving one or two spins.  This local update is performed consecutively in a sweeping fashion through the lattice.  This choice yields large autocorrelation times, as many updates need to be performed to construct a sample $\mathcal{S}^f$ that is statistically independent from some initial sample $\mathcal{S}^i$.  Therefore, while performing the Markov Chain, one only adds samples to the sample set periodically, with a time interval $\tau_s$ that has to be chosen greater than the autocorrelation time $\tau_a$ defined by the decay of correlations of the relevant operators at different times
\begin{multline}
    \sum_{t=0}^{\infty} \langle O(\mathcal{S}^{t+t'}) O(\mathcal{S}^t) \rangle - \langle O(\mathcal{S}^{t+t'}) \rangle \langle O(\mathcal{S}^t) \rangle \\ \sim \exp ( -t' / \tau_a ).
\end{multline}
Another disadvantage of MCMC algorithms is that the Markov Chain is started from an initial configuration $\mathcal{S}^i$, that is typically not drawn according to the distribution $|\Psi(\mathcal{S})|^2$.  Starting from $\mathcal{S}^i$, consecutive samples are only distributed according to $|\Psi(\mathcal{S})|^2$ in the limit of infinite time.  In practice, one performs an equilibration period $\tau_i$ that is sufficiently long such that the value of an observable becomes approximately stationary when estimated using different Markov Chains. The period $\tau_i$ is determined by the gap between the largest and second largest eigenvalues of $T(\mathcal{S}^1 \rightarrow \mathcal{S}^2)$, and the samples drawn during this equilibration period are not usable for estimating expectation values. Determining $\tau_i$ and $\tau_s$ is computationally hard and, as such, heuristics are used to determine them in practice.

The choice for local updates in VMC algorithms is imposed by the fact that during every optimization step, a different probability distribution $\left|\Psi(\mathcal{S})\right|^2$ is encountered.  Therefore, it is impractical to determine a rule $\omega(\mathcal{S}^1 \rightarrow \mathcal{S}^2)$ that is optimal at every optimization step, as this requires much knowledge about the structure of the probability distribution.  When this knowledge does exist -- e.g. for classical Boltzmann distributions with nearest-neighbor interactions -- highly efficient trial probabilities can be constructed.  An example is the Wolff algorithm~\cite{Wolff1989Collective} performing updates of whole clusters.

\subsection{Direct sampling}
\label{sec:ds}

In order to alleviate these shortcomings, we introduce a method for drawing independent samples from a PEPS tensor network, also called direct sampling (DS).  In order to have an efficient direct-sampling scheme, we (i) directly sample from an effective distribution $p_c(\S)$ that is not too far from $p_\Psi(\S)$, (ii) provide a way to compute the ratio $|\Psi(\mathcal{S})|^2/p_c(\S)$ for each sample that we draw and (iii) use this ratio in an importance sampling algorithm to determine expectation values. The distribution $p_c$ will be constructed as a sequence of conditional probabilities for each row.
\par As explained in Sec.~\ref{sec:peps}, we can view the contraction of the double-layer network as the evolution of a one-dimensional density matrix. We define
\begin{align}
\rho_1 = 
\begin{array}{c}
\begin{tikzpicture}[scale=.5]
\draw[shift={(0+3/4,0+3/4)}] (0,0)--(6,0);
\foreach \x in {0,2,...,6}{
\draw[shift={(\x+3/4,0+3/4)}] (0,0)--(0,-1);
\draw[color=white,myBG,shift={(\x,0)}] (0,0) -- (0,-1);
\draw[shift={(\x,0)}] (0,0)--(0,-1);
\draw[color=white,myBG,shift={(0,0)}] (0,0) -- (6,0);
\draw[shift={(0,0)}] (0,0) -- (6,0);
};
\foreach \x in {0,2,...,6}{
    \foreach \y in {0}{
    	\filldraw[lightten,shift={(\x+3/4,\y+3/4)}]  (-1/2,-1/2)--(1/2,-1/2)--(1/2,1/2)--(-1/2,1/2)--(-1/2,-1/2);    	\draw[shift={(\x,\y)}]  (0,0)--(3/4,3/4);
    	\filldraw[tenred,shift={(\x+2/4,\y+2/4)}] (0,0) circle (0.150);
    	\filldraw[ten,shift={(\x,\y)}]  (-1/2,-1/2)--(1/2,-1/2)--(1/2,1/2)--(-1/2,1/2)--(-1/2,-1/2);
	};
};
\end{tikzpicture}
\end{array},
\end{align}
and 
\begin{align}
\rho_2 = 
\begin{array}{c}
\begin{tikzpicture}[scale=.5]
\draw[shift={(0+3/4,0+3/4)}] (0,0)--(6,0);
\draw[shift={(0+3/4,-2+3/4)}] (0,0)--(6,0);
\foreach \x in {0,2,...,6}{
\draw[shift={(\x+3/4,0+3/4)}] (0,0)--(0,-3);
\draw[color=white,myBG,shift={(\x,0)}] (0,0) -- (0,-3);
\draw[shift={(\x,0)}] (0,0)--(0,-3);
};
\draw[color=white,myBG,shift={(0,0)}] (0,0) -- (6,0);
\draw[shift={(0,0)}] (0,0) -- (6,0);
\draw[color=white,myBG,shift={(0,-2)}] (0,0) -- (6,0);
\draw[shift={(0,-2)}] (0,0) -- (6,0);
\foreach \x in {0,2,...,6}{
    \foreach \y in {0,-2}{
    	\filldraw[lightten,shift={(\x+3/4,\y+3/4)}]  (-1/2,-1/2)--(1/2,-1/2)--(1/2,1/2)--(-1/2,1/2)--(-1/2,-1/2);    	\draw[shift={(\x,\y)}]  (0,0)--(3/4,3/4);
    	\filldraw[tenred,shift={(\x+2/4,\y+2/4)}] (0,0) circle (0.150);
    	\filldraw[ten,shift={(\x,\y)}]  (-1/2,-1/2)--(1/2,-1/2)--(1/2,1/2)--(-1/2,1/2)--(-1/2,-1/2);
	};
};
\end{tikzpicture}
\end{array},
\end{align}
where the uncontracted bonds denote a density matrix defined on the vector space of the boundary.
The evolution of the density matrix can be written as
\begin{equation}
    \rho_{y} =  \sum_{\{\mathcal{S}_{y}\}} c_y({\mathcal{S}_{y}}) \rho_{y-1} c_y({\mathcal{S}_{y}})^{\dag},
    \label{eq:master_peps}
\end{equation}
where $c_y(\mathcal{S}_y)$ is the $y$-th row of the PEPS projected on the physical basis states $\mathcal{S}_y$.
This equation can be seen as a discrete version of the Lindblad equation
\begin{equation}
    \frac{d \rho}{dt} =  - i (H_{\mathrm{eff}} \rho -\rho H_{\mathrm{eff}}^{\dag}) + \sum_m c_m \rho c_m^{\dag},
    \label{eq:master}
\end{equation}
where we identify the effective Hamiltonian $H_{\mathrm{eff}} = -\frac{i}{2} \mathbb{1}$ and the Lindblad operators $c_m = c_y({\mathcal{S}_{y}})$. One approach for integrating the Lindblad equation consists of stochastically obtaining a set of trajectories and estimating expectation values from this set. The trajectories consist of time evolutions of pure states, where the dynamics according to $H_{\mathrm{eff}}$ is alternated with applications of the operators $c_m$, the so-called quantum jumps. The full algorithm is detailed in \cite{Daley2014QuantumSystems} and an overview can be found in Appendix~\ref{sec:app_QJMC}. 
\par For our purpose of sampling PEPS double-layer networks, we can use this quantum jump algorithm to simulate trajectories that can be identified by configurations $\mathcal{S}$. We start from the trivial boundary density operator with dimension one
\begin{align}
\rho_0 = \mathbb{1} =
\begin{array}{c}
\begin{tikzpicture}[scale=.5]
\foreach \x in {0,2,...,6}{
    \foreach \y in {0}{
    	\draw (\x+3/8,0) to[out=0, in=90, distance=0.25cm] (\x+3/4,-1+1/4);
    \draw (\x+3/8,0) to[out=0, in=-90, distance=-0.5cm] (\x,-1-1/4);
	};
};
\end{tikzpicture}
\end{array}.
\label{eq:trivial_boundary}
\end{align}
Locally, this density operator can be written as the outer product of a one-dimensional unit vector with itself.  For the full boundary, this yields $\rho_0 = \ket{\phi_{0}}_b \bra{\phi_{0}}_b$, with 
\begin{align}
\ket{\phi_{0}}_b = 
\begin{array}{c}
\begin{tikzpicture}[scale=.5]
\foreach \x in {0,2,...,6}{
\draw[color=white,myBG,shift={(\x,0)}] (0,0) -- (0,-1);
\draw[shift={(\x,0)}] (0,0)--(0,-1);
};
\foreach \x in {0,2,...,6}{
    \foreach \y in {0}{
 	    \filldraw[black,shift={(\x,\y)}] (0,0) circle (0.150);
    };
};
\end{tikzpicture}
\end{array},
\end{align}
where the subscript denotes that the bra and ket pertain to the vector space of the boundary, and not the physical space.  The boundary vector $\ket{\phi_{0}}_b$ can be written in MPS form with bond dimension one
\begin{align}
\ket{\phi_{0}}_b = 
\begin{array}{c}
\begin{tikzpicture}[scale=.5]
\foreach \x in {0,2,...,6}{
\draw[color=white,myBG,shift={(\x,0)}] (0,0) -- (0,-1);
\draw[shift={(\x,0)}] (0,0)--(0,-1);
};
\draw[color=white,myBG,shift={(0,0)}] (0,0) -- (6,0);
\draw[shift={(0,0)}] (0,0) -- (6,0);
\foreach \x in {0,2,...,6}{
    \foreach \y in {0}{
 	    \filldraw[black,shift={(\x,\y)}] (0,0) circle (0.150);
    };
};
\end{tikzpicture}
\end{array}.
\end{align}

Starting from this trivial boundary MPS, we propose a sequence of three steps for moving down one row $y$ in the single-layer PEPS, finding a sample $\S_y$ for the row's configuration space as we go:
\begin{enumerate}
    \item Apply a row of the single-layer PEPS tensors to the boundary MPS from the previous step,
    \begin{align}
    \mathbf{c_y}\ket{\phi_{y-1}}_b &= 
    \begin{array}{c}
    \begin{tikzpicture}[scale=.5]
    \foreach \x in {0,2,...,6}{
    \draw[color=white,myBG,shift={(\x,0)}] (0,1) -- (0,-1);
    \draw[shift={(\x,0)}] (0,1)--(0,-1);
    };
    \draw[color=white,myBG,shift={(0,0)}] (0,0) -- (6,0);
    \draw[shift={(0,0)}] (0,0) -- (6,0);
    \draw[color=white,myBG,shift={(0,1)}] (0,0) -- (6,0);
    \draw[shift={(0,1)}] (0,0) -- (6,0);
    \foreach \x in {0,2,...,6}{
        \foreach \y in {0}{
     	    \draw[shift={(\x,\y)}]  (0,0)--(3/4,3/4);
     	    \filldraw[black,shift={(\x,\y+1)}] (0,0) circle (0.150);
    	    \filldraw[ten,shift={(\x,\y)}]  (-1/2,-1/2)--(1/2,-1/2)--(1/2,1/2)--(-1/2,1/2)--(-1/2,-1/2);
	    };
    };
    \end{tikzpicture}
    \end{array} \nonumber \\
    &=
    \begin{array}{c}
    \begin{tikzpicture}[scale=.5]
    \foreach \x in {0,2,...,6}{
    \draw[color=white,myBG,shift={(\x,0)}] (0,0) -- (0,-1);
    \draw[shift={(\x,0)}] (0,0)--(0,-1);
    };
    \draw[color=white,myBG,shift={(0,0)}] (0,0) -- (6,0);
    \draw[shift={(0,0)},line width=2] (0,0) -- (6,0);
    \foreach \x in {0,2,...,6}{
        \foreach \y in {0}{
     	    \draw[shift={(\x,\y)}]  (0,0)--(3/4,3/4);
    	    \filldraw[tenorange,shift={(\x,\y)}]  (-1/2,-1/2)--(1/2,-1/2)--(1/2,1/2)--(-1/2,1/2)--(-1/2,-1/2);
	    };
    };
    \end{tikzpicture}
    \end{array},
    \end{align}
    where we introduce the tensor $\mathbf{c_y}$ as the collection of all operators $c_y(\mathcal{S}_y)$, i.e. it is an MPS-like state, where every tensor has an extra index originating from the physical degrees of freedom. After obtaining $\mathbf{c_y}\ket{\phi_{y-1}}_b$, the bond dimension of the boundary has grown, and it should be truncated if necessary. In general, we will impose a maximal bond dimension on $\mathbf{c_y}\ket{\phi_{y-1}}_b$, which we will denote $\chi_s$.
    \item Sample one of the configurations $\mathcal{S}_y$ according to the probability distribution
    \begin{multline} \label{eq:pc}
        \hspace{1cm} p_c(\mathcal{S}_y|\mathcal{S}_{< y}) \\ = \frac{\braket{\phi_{y-1}|_b c(\mathcal{S}_y)^{\dag} c(\mathcal{S}_y)|\phi_{y-1}}_b}
        {\sum_{\{\mathcal{S}_y\}}\braket{\phi_{y-1}|_b c(\mathcal{S}_y)^{\dag} c(\mathcal{S}_y)|\phi_{y-1}}_b}.
    \end{multline}
    This sampling step is equivalent to the sampling of an MPS and can be efficiently performed by using the perfect-sampling approach for unitary tensor networks \cite{Ferris2012PerfectNetworks}. For convenience, we have summarized this algorithm in Appendix~\ref{density_sampling}.
    \item When a configuration $\mathcal{S}_y = (\mathcal{S}_{1,y},...,\mathcal{S}_{L_x,y})$ is found, one obtains a new boundary MPS $\ket{\phi_y}_b$ by projecting the physical degrees of freedom in $\mathbf{c_y}\ket{\phi_{y-1}}_b$ onto the configuration $\mathcal{S}_y$,
    \begin{align}
    \ket{\phi_y}_b &= 
    \begin{array}{c}
    \begin{tikzpicture}[scale=.5]
    \foreach \x in {0,2,...,6}{
    \draw[color=white,myBG,shift={(\x,0)}] (0,0) -- (0,-1);
    \draw[shift={(\x,0)}] (0,0)--(0,-1);
    };
    \draw[color=white,myBG,shift={(0,0)}] (0,0) -- (6,0);
    \draw[shift={(0,0)},line width=2] (0,0) -- (6,0);
    \foreach \x in {0,2,...,6}{
        \foreach \y in {0}{
     	    \draw[shift={(\x,\y)}]  (0,0)--(1,1.3);
     	    \filldraw[tenwhite,shift={(\x+1,\y+1.3)}] (0,0) circle (0.70);
    	    \filldraw[tenorange,shift={(\x,\y)}]  (-1/2,-1/2)--(1/2,-1/2)--(1/2,1/2)--(-1/2,1/2)--(-1/2,-1/2);
	    };
    };
    \node at(0+1,1.3) {$\mathcal{S}_{1,y}$};    
    \node at(2+1,1.3) {$\mathcal{S}_{2,y}$};
    \node at(4+1,1.3) {$\mathcal{S}_{3,y}$};
    \node at(6+1,1.3) {$\mathcal{S}_{4,y}$};
    \end{tikzpicture}
    \end{array} \nonumber \\
    &=
    \begin{array}{c}
    \begin{tikzpicture}[scale=.5]
    \foreach \x in {0,2,...,6}{
    \draw[color=white,myBG,shift={(\x+1/2,0)}] (0,0) -- (0,-1);
    \draw[shift={(\x+1/2,0)}] (0,0)--(0,-1);
    };
    \draw[color=white,myBG,shift={(0,0)}] (0,0) -- (6,0);
    \draw[shift={(1/2,0)}] (0,0) -- (6,0);
    \foreach \x in {0,2,...,6}{
        \foreach \y in {0}{
    	    \filldraw[black,shift={(\x,\y)}] (1/2,0) circle (0.150); (-1/2,-1/2)--(1/2,-1/2)--(1/2,1/2)--(-1/2,1/2)--(-1/2,-1/2);
	    };
    };
    \end{tikzpicture}
    \end{array}.
    \label{eq:result_MPS}
    \end{align}
    For convenience, we divide this boundary MPS by $\sqrt{p_c(\mathcal{S}_y|\mathcal{S}_{< y})}$.
\end{enumerate}
After going through all $L_y$ rows of the PEPS, we obtain a sample configuration $\mathcal{S}$, sampled with the sequence of conditional probabilities
\begin{equation}
    p_{c}(\mathcal{S}) = p_c(\mathcal{S}_1)p_c(\mathcal{S}_2|\mathcal{S}_1) \; \dots\; p_c(\mathcal{S}_{L_y}|\mathcal{S}_{<L_y}).
    \label{eq:def_pc}
\end{equation}
This equation highlights the viewpoint that $p_{c}(\mathcal{S})$ is built as a sequence of conditional probability distributions of row configurations.  As such, the algorithm follows the same strategy as autoregressive sampling techniques and also the direct sampling algorithm for one dimensional loop-free tensor networks \cite{Oord2016WaveNet, pmlr-v48-oord16, Ferris2012PerfectNetworks}.  In these algorithms, one assumes that marginal probability distributions can be efficiently calculated or have a specific structure by construction of the model.  Note that we arrived at Eq.~\eqref{eq:def_pc} by assuming that the marginals of $p_{\Psi}(\mathcal{S})$ are equal to the unit reduced density matrix.  Hence, $p_c(\mathcal{S})$ is related but not equal to $p_{\Psi}(\mathcal{S})$.
Indeed, the prescription for $p_c$ in Eq.~\eqref{eq:pc} is a drastic approximation for the true marginal probability distributions of $p_{\Psi}(\mathcal{S})$ for the variables $\S_y$, as it ignores all other rows in the network. We can systematically improve on the above scheme such that the distribution $p_{c}(\mathcal{S})$ comes closer to $p_{\Psi}(\mathcal{S})$. Indeed, we can improve the approximation in step 2 of the algorithm by tracing out the environment of the trailing rows approximately, 
\begin{align}
E_m(\mathcal{S}_{>y}) &= 
\begin{array}{c}
\begin{tikzpicture}[scale=.5]
\draw[shift={(0+3/4,0+3/4)}] (0,0)--(6,0);
\draw[shift={(0+3/4,2+3/4)}] (0,0)--(6,0);
\foreach \x in {0,2,4,6}{
\draw[shift={(\x+3/4,0+3/4)}] (0,0)--(0,3);
\draw[color=white,myBG,shift={(\x,0)}] (0,0) -- (0,3);
\draw[shift={(\x,0)}] (0,0) -- (0,3);
};
\draw[color=white,myBG,shift={(0,0)}] (0,0) -- (6,0);
\draw[color=white,myBG,shift={(0,2)}] (0,0) -- (6,0);
\draw[shift={(0,0)}] (0,0)--(6,0);
\draw[shift={(0,2)}] (0,0)--(6,0);
\foreach \x in {0,2,4,6}{
    \foreach \y in {0,2}{
    	\filldraw[lightten,shift={(\x+3/4,\y+3/4)}]  (-1/2,-1/2)--(1/2,-1/2)--(1/2,1/2)--(-1/2,1/2)--(-1/2,-1/2);    	\draw[shift={(\x,\y)}]  (0,0)--(3/4,3/4);
    	\filldraw[ten,shift={(\x,\y)}]  (-1/2,-1/2)--(1/2,-1/2)--(1/2,1/2)--(-1/2,1/2)--(-1/2,-1/2);
	};
};
\end{tikzpicture}
\end{array} \nonumber \\
&= \begin{array}{c}
\begin{tikzpicture}[scale=.5]
\draw[color=white,myBG,shift={(0,0)}] (0,0) -- (6,0);
\draw[shift={(0,0)},line width=2] (0,0)--(6,0);
\foreach \x in {0,2,4,6}{
    \foreach \y in {0}{
        \draw (\x,\y) to[out=25, in=-90, distance=0.5cm] (\x+3/4,\y+1);
    	\filldraw[tengreen,shift={(\x,\y)}]  (-1/2,-1/2)--(1/2,-1/2)--(1/2,1/2)--(-1/2,1/2)--(-1/2,-1/2);
    	\draw (\x,\y) to[out=-25, in=90, distance=-0.5cm] (\x-3/4,\y+1);
	};
};
\end{tikzpicture}
\end{array}.
\label{eq:Em}
\end{align}
These environments are MPOs, that can be truncated to a maximal bond dimension $\chi_m$. In the sampling algorithm, using these environments boils down to inserting this MPO in the above prescription for the marginal distribution $p_c$
\begin{multline} \label{eq:pcm}
p_c^m(\mathcal{S}_y|\mathcal{S}_{<y}) \\ = \frac{\braket{\phi_{y-1} |_b c(\mathcal{S}_y)^{\dag} E_m(\mathcal{S}_{>y})  c(\mathcal{S}_y) |\phi_{y-1}}_b} {\sum_{\{\mathcal{S}_y\}}\braket{\phi_{y-1}|_b c(\mathcal{S}_y)^{\dag} E_m(\mathcal{S}_{>y}) c(\mathcal{S}_y)|\phi_{y-1}}_b}
\end{multline}
By increasing the environment bond dimension $\chi_m$, the distribution $p_c^m(\S)$ that is sampled in our scheme will converge to $p_\Psi(\S)$. Finding the environments $E_m$, however, requires a boundary-MPO contraction of the double-layer network, which we were set out to avoid in the first place. We will show in the following section that using small values of $\chi_m$ yield environments that are computationally cheap to compute, compared to the rest of the algorithm, and already yield satisfactory results -- although a full double-layer computation with this value of the boundary-MPO bond dimension would not yield an acceptable accuracy.

The expectation value in Eq.~\eqref{eq:expectation} can now be written as
\begin{align}
    \langle \widehat{O} \rangle &= \frac{1}{\braket{\Psi|\Psi}}\sum_{\{\mathcal{S}\}} p_c(\mathcal{S}) \frac{|\Psi(\mathcal{S})|^2}{p_c(\mathcal{S})} O_{\loc}(\mathcal{S}) \nonumber \\
    &\approx \frac{1}{N\braket{\Psi|\Psi}} \sum_{\mathcal{S} \sim p_{c}} \frac{|\Psi(\mathcal{S})|^2}{p_c(\mathcal{S})} O_{\loc}(\mathcal{S}).
    \label{eq:qj_expectation}
\end{align}
In this equation, the term $|\Psi(\mathcal{S})|^2/p_c(\mathcal{S})$ can be readily computed after a sample is drawn by calculating the norm of Eq.~\eqref{eq:result_MPS} as $\ket{\phi_y}_b$ is essentially a single-layer contraction $\Psi(\mathcal{S})$ divided by the root of $p_c(\mathcal{S})$ as defined in Eq.~\eqref{eq:def_pc}. The norm $\braket{\Psi|\Psi}$ can also be computed by using the resulting samples as it can be written as
\begin{equation}
    \braket{\Psi|\Psi} = \sum_{\{\mathcal{S}\}} p_c(\mathcal{S}) \frac{|\Psi(\mathcal{S})|^2}{p_c(\mathcal{S})}
    \approx \frac{1}{N} \sum_{\mathcal{S} \sim p_c} \frac{|\Psi(\mathcal{S})|^2}{p_c(\mathcal{S})}.
    \label{eq:norm}
\end{equation}
Eq.~\eqref{eq:qj_expectation} is expressed as an importance sampling procedure, sampling from $p_c(\mathcal{S})$.  Hence, the estimate of the expectation value is asymptotically unbiased.
The variance on the expectation value in Eq.~\eqref{eq:qj_expectation} is defined as
\begin{multline}
    \textrm{Var}_{p_c}(\langle \widehat{O} \rangle) = \\ 
    \left< \left( \frac{1}{N} \sum_{\mathcal{S} \sim p_c} \left( \frac{|\Psi(\mathcal{S})|^2  O_{loc}(\mathcal{S})}{\braket{\Psi|\Psi} p_c(\mathcal{S})} - \langle \widehat{O} \rangle \right) \right)^2 \right>,
    \label{eq:var_pc}
\end{multline}
i.e. the expected square deviation of the estimate of the observable using $N$ samples from the true observable.

It is clear that the disadvantages related to autocorrelation in a Markov Chain do not arise in this algorithm. The samples are drawn independently from each other and no equilibration period is needed. A second advantage is the possibility to estimate the norm of a PEPS as shown in Eq.~\eqref{eq:norm}, which is not possible when sampling from $p_{\Psi}$ by a Markov chain. Furthermore, as the algorithm has no interdependence between samples, it can be executed in a massively parallel fashion.

\subsection{Independent samples as Markov Chain transitions}

Finally we can combine both sampling approaches. Indeed, we can take the samples drawn according to $p_c$ as the transition probabilities in a Markov Chain. Concretely, we take as transition probabilities $\omega(\mathcal{S}^1 \rightarrow \mathcal{S}^2) \equiv \omega(\mathcal{S}^2) = p_c(\mathcal{S}^2)$.  In this case, the proposed configuration is independent from the previous one, and the new configuration is accepted with the acceptance probability defined in Eq.~\eqref{eq:acceptance}. As detailed balance (Eq.~\eqref{eq:detailed_balance}) is satisfied, the resulting distribution from which the samples are drawn is $p_{\Psi}$ . This algorithm still has the advantage of drawing independent samples and being massively parallelizable.  The downside is that an equilibration period is needed because, at the start of this Markov Chain, the samples are drawn according to $p_c$, which is not equal to $p_{\Psi}$.  This approach is similar to that of Ref.~\cite{Frias-Perez2021CollectiveRenormalization}, where the transition probabilities $T(\mathcal{S}^2)$ were determined by approximately contracting a single layer tensor network representing a classical Boltzmann distribution.

\section{Stochastic optimization of PEPS}
\label{sec:optimization}

Given that we can efficiently sample a PEPS wavefunction and compute expectation values, we now give details on how to variationally optimize the PEPS tensors with respect to a given model Hamiltonian.
\par The first issue concerns the initialization of the PEPS tensors. In recent works, an initial state was constructed by running the simple-update algorithm for PEPS, stating that starting from an initial PEPS with randomly initialized parameters leads to poor statistics and local minima~\cite{AccurateLiu2021}. In this work, however, we have observed that this additional step is unnecessary by choosing a PEPS of which the weights are randomly initialized from a uniform distribution between zero and one. With this initialization, the distribution of the expansion coefficients $\Psi(\mathcal{S})$ is narrowly distributed around a non-zero value.  This is qualitatively dissimilar from the distribution obtained with a PEPS whose weights are initialized with positive and negative weights. Indeed, as a single-layer PEPS contraction can be viewed as a sum of products of tensor elements, this sum will have many canceling contributions, leading to a distribution centered and peaked around zero.  When considering local operators $O_{\textrm{loc}}(\mathcal{S})$, the expansion coefficients enter in the denominator and expansion coefficients close to zero lead to large contributions to the local operator.  Note that, with a finite sample set, the large contributions lead to poor statistics of the expectation value.  On the other hand, when the expansion coefficients are narrowly distributed around a non-zero value, the matrix elements of the operator $\braket{\mathcal{S}|\widehat{O}|\mathcal{S}'}$ have a larger effect on the value of the local operator and the expectation value has low variance.  
We also rescale the weights by properly normalizing the PEPS, ensuring that the mean of the distribution $|\Psi(\mathcal{S})|^2/p_c(\mathcal{S})$ is equal to one.  

It is often worthwhile to consider symmetries of the models under study when optimizing variational ground states. In this work, when present, we will consider eigenstates of the $U(1)$ symmetry group by projecting the PEPS on the subspace with total spin projection equal to a fixed value $Q$
\begin{equation}
    \ket{\Psi_{S^z=Q}} = \widehat{P}_{S^z=Q} \ket{\Psi} , \qquad S^z=\sum_{x,y} S^z_{x,y}.
\end{equation}
This projector can be efficiently implemented by sampling only configurations $\mathcal{S}$ with the constraint $\bra{\S} \hat{S}_z \ket{\S}=Q$. Indeed, we start the sampling as is described above and we keep track of the total $U(1)$ charge accumulated during the sampling. At some point, we reach a bound that is determined by the desired value $Q$ and the total number of spins; from that point on, we define the distributions $p_c(\mathcal{S}_{x,y}|\mathcal{S}_{<x,<y})$ such that it is zero for $\mathcal{S}_{x,y}$ equal to the species that reached its bound.  In this way, we prohibit the amount of spins of a species to grow beyond its bound.
\par Note that this approach for imposing a $U(1)$ symmetry (or any abelian symmetry) is different than the standard one in the tensor-network formalism, where a global (in general, non-abelian) symmetry of a given tensor-network state can be encoded as a constraint on the level of the local tensors. Here, we apply a global projector on the full state by constraining the sampling procedure. In a direct contraction, applying this projector would require a prohibitively large bond dimension.

After initialization, the parameters of the PEPS are updated using an estimate of the gradient of the expectation value of the Hamiltonian with respect to the tensors defined in Eq.~\eqref{eq:tensor}. The expression for these gradients is
\begin{multline}
    g = \frac{\partial \langle \widehat{H} \rangle}{\partial A^{(x,y)}_{\alpha \beta \gamma \delta}(s)} \\
    = \sum_{\{\mathcal{S}\}} \frac{|\Psi(\mathcal{S})|^2}{\braket{\Psi|\Psi}} \frac{\partial \ln(\Psi^*(\mathcal{S}))}{\partial A^{(x,y)}_{\alpha \beta \gamma \delta}(s)} \left( H_{\loc}(\mathcal{S}) - \langle \widehat{H} \rangle \right) ,
    \label{eq:gradient}
\end{multline}
where we denote the coordinates of the different tensors as $(x,y)$. The gradient provides the direction which maximizes the expectation value of the Hamiltonian locally in parameter space; by updating the parameters in the opposite direction using a small step size, we move the PEPS state closer to the minimum of the energy, which is known as the stochastic gradient-descent (SGD) algorithm.  Note that the gradient as calculated by sampling, and more generally any expectation value, is inherently noisy due to the finite amount of samples.  The noise will have an effect on the gradient direction, i.e. not pointing exactly in the opposite direction of steepest descent.  However, the accumulation of different noisy steps will yield a trajectory in parameter space that moves in the direction of lower energies because the average direction of different noisy gradients is still close to the direction of steepest descent, and thus a descent direction.  Furthermore, the level of noise can be controlled by taking more samples into account in the estimation of expectation values.

To alleviate this problem, one can use information from previous steps.  An example is the Adam optimization algorithm \cite{kingma2014adam}, that uses an exponentially weighted average of previous gradients to construct the direction of the current step.  This is combined with a parameter-dependent step size, scaling as the exponentially weighted average of the norm of the gradient of a parameter.

Furthermore, we can refine the gradient descent approach by including information about the shape of the manifold described by the PEPS ansatz and proposing an update direction based on this information combined with the gradient. In the VMC context, this is the so-called stochastic reconfiguration (SR) approach \cite{Sorella1998GreenReconfiguration,Neuscamman2012OptimizingCarlo}, also known as natural gradient descent \cite{Amari1998NaturalLearning}. An alternative viewpoint is that it includes information about the metric of the PEPS manifold or, equivalently, performs an imaginary time evolution step on the PEPS manifold according to the time-dependent variational principle \cite{Haegeman2016UnifyingStates, Haegeman2011Time-dependentLattices, Vanderstraeten2019tangent}.  In any case, we find a more natural descent direction by solving the following linear equation for the direction $\mathbf{x}$
\begin{equation}
    \mathbf{S} \mathbf{x} = \mathbf{g},
    \label{eq:SR}
\end{equation}
where we define the matrix $\mathbf{S}$ as
\begin{multline}
    \mathbf{S}_{i,j} = \left< \frac{\partial \ln(\Psi^*)}{\partial p_i} \frac{\partial \ln(\Psi)}{\partial p_j} \right> \\
    - \left< \frac{\partial \ln(\Psi^*)}{\partial p_i} \right> \left< \frac{\partial \ln(\Psi)}{\partial p_j} \right>.
\end{multline}
Here the indices $i$ and $j$ denote collective indices for all variational parameters in the PEPS wavefunction. For a given sample set, the matrix $\mathbf{S}$ can be singular and hence the inversion needed to solve Eq.~\eqref{eq:SR} for $\mathbf{x}$ can fail.  In practical applications described below, we will regularize the matrix $\mathbf{S}$ by shifting the diagonal of the matrix with an amount $\delta$.  In all simulations below, we use $\delta=0.01$.  Solving the linear system of equations in Eq.~\eqref{eq:SR} can be done with an iterative solver such as the conjugate gradient method.  Using this algorithm, one has to perform the multiplication of $\mathbf{S}$ with some arbitrary vector, which can be performed efficiently by sampling $\mathbf{S}$, scaling with the number of variational parameters, i.e. $\mathcal{O}(D^4)$.

\section{Results}
\label{sec:results}

\subsection{Performance benchmark}

\begin{figure}
\includegraphics{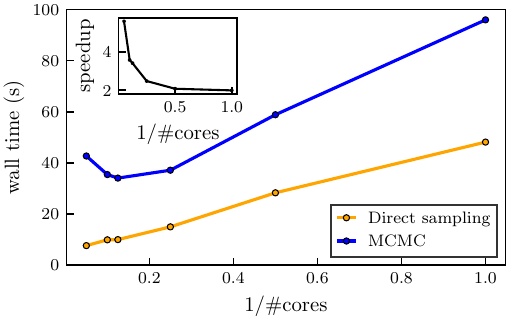}
\caption{Comparison of wall time as a function of the inverse of the number of CPU cores used for direct sampling and MCMC.  Every point is the median of 10 separate runs.  Every run consists of drawing $N=1000$ samples, where an equilibartion time of $\tau_i=100$ sweeps was used for the MCMC runs. The inset shows the relative speedup of the direct sampling procedure compared to MCMC for constructing a full set of samples.}
  \label{fig:time_benchmark}
\end{figure}
As noted in Sec.~\ref{sec:ds}, direct sampling aims to solve downsides of MCMC algorithms such as the problem of autocorrelation and equilibration.  The problem of equilibration translates to the fact that MCMC algorithms can not be parallelized indefinitely, because the sampling during the equilibration period is inherently serial.  In Fig.~\ref{fig:time_benchmark}, we show the wall time as a function of the number of CPU cores for drawing a sample set of $N=1000$ samples from a $10 \times 10$ PEPS with bond dimension $D=4$.  We compare direct sampling with MCMC, both using a cut-off $\chi_s=4$ in the contractions.  For MCMC, we use an equilibration period of $\tau_i=100$ sweeps and $\tau_s=1$ sweep between consecutive samples.  We define a sweep as performing a local update once on every site of the lattice, visiting the lattice sites row by row, going from left to right within every row.   For direct sampling, we use a cut-off $\chi_m=4$ for the environments of the marginals.  In theory, the wall time of direct sampling scales as $t=\mu / N_c$ and that of MCMC scales as $t=\nu \tau_i + \xi \tau_s / N_c$, where $N_c$ is the number of cores used.  In practice, computational overhead increases the runtime on an actual computer.  It is clear that direct sampling is twice as fast for a low number of CPU cores, while it progressively becomes more efficient when using more CPU cores, reaching a five-fold speedup for $20$ CPU cores.  The MCMC algorithm suffers from the combination of numerical overhead and the equilibration when using more than $10$ CPU cores, while overhead is much less severe in the case of direct sampling.

\subsection{Benchmark on classical spin models}

As a proof of principle, we apply our sampling algorithm to drawing samples according to the partition function of the classical two-dimensional Ising model. The classical Ising model is defined by the energy function
\begin{equation}
    E(\S) = -J \sum_{\langle ij,kl \rangle} \S_{i,j} \S_{k,l},
\end{equation}
where $\S_{i,j}$ are classical spin variables and the sum iterates over all pairs of nearest neighbors. The partition function can be written as the norm of a PEPS wavefunction \cite{Verstraete2006Criticality}. Indeed, by defining the tensors
\begin{equation}
A_{\alpha \beta \gamma \delta}(s) = 
\begin{cases}
1, & \textrm{if}\ \alpha=\beta=\gamma=\delta=s \\
0, & \textrm{otherwise},
\end{cases}
\end{equation}
and
\begin{equation}
B = 
\begin{bmatrix}
e^{1/2kT} & e^{-1/2kT} \\
e^{-1/2kT} & e^{1/2kT} 
\end{bmatrix},
\end{equation}
and arranging the tensors $A$ on the lattice sites of the square lattice and the tensors $B$ on the links between the lattice sites, one sees that the resulting PEPS state represents the vector of square root probabilities of the Ising model.  In the following, we compare our sampling algorithm with the Metropolis-Hastings algorithm using single spin flips, and the Wolff algorithm \cite{Wolff1989Collective}. The former is a very versatile variant of MCMC, which is also typically used for the optimization of quantum states.  The latter is specifically designed for Ising-like models, and is well-known for its favorable scaling of the (integrated) autocorrelation time.  

\begin{figure}
\includegraphics{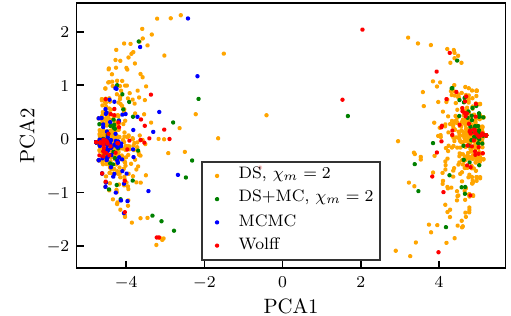}
\caption{Projection of data points using different sampling approaches on two principal components, for a PEPS encoding the classical Ising model.  The principal components are those with largest explained variance of the correlation matrix of the sample set constructed with direct sampling.  The first component coincides with magnetization of the samples.  MCMC, which is in this case equivalent to the Metroplois-Hastings algorithm applied to the Boltzmann distribution, is clearly unable to sample both modes of the probability distribution.  The Wolff algorithm and the direct sampling approaches do not suffer from this problem.}
  \label{fig:pca}
\end{figure}
In Fig.~\ref{fig:pca} we show a principal component analysis of sampled configurations of a $10\times10$ Ising model in the ferromagnetic phase.  Specifically, we show the two components with largest explained variance, i.e. the projection of the data on the two eigenvectors with largest eigenvalue of the covariance matrix of the data~\cite{Pearson1901}. For every algorithm, we sample 10000 configurations and subsample with equal intervals 1000 samples of our PEPS sampling algorithm and 100 samples of the other algorithms.  The first principal component is closely related to the magnetization, and shows the bimodal distribution of it.  Our sampling algorithm is symmetric with respect to the sign of the magnetization, unlike the MCMC algorithm which tends to be unable to cross the large energy barrier between the two modes.  Also when using our sampling algorithm as transitions in a Markov chain (denoted DS+MC), the resulting samples have equal weight in both modes of the magnetization, as the transitions are independent of the previous samples.  This shows that our sampling algorithm can be used to cross large barriers in probability space, as is necessary in, e.g, glassy models~\cite{Binder1986, Swendsen1986, PhysRevE.104.034105}.

\begin{figure}
\includegraphics{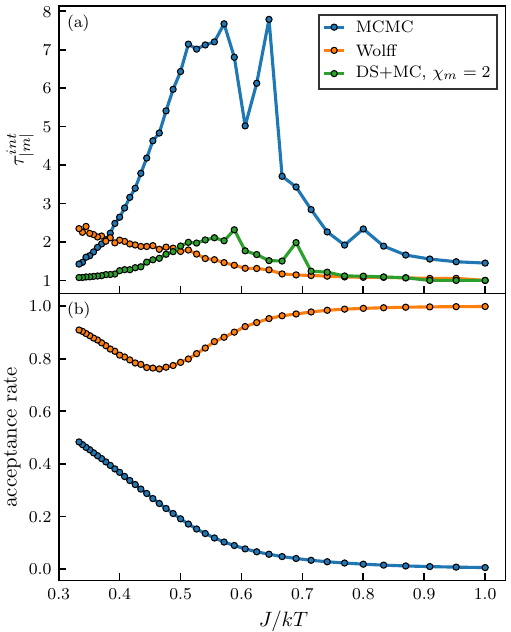}
\caption{Comparison of sampling efficiency of a $10 \times 10$ classical square-lattice Ising model between MCMC, Wolff and our direct sampling + MC approach. For the direct sampling + MC approach, we use $\chi_m=D=2$. (a), The integrated autocorrelation time as defined in Eq.~\eqref{eq:tau_int} as a function of $J/kT$.  MCMC shows clear slowing down in the vicinity of the critical point, while this effect is much less severe in the case of direct sampling + MC and Wolff.  (b), Acceptance rate of the Markov chain used in MCMC and direct sampling + MC.}
  \label{fig:ac}
\end{figure}
As a second benchmark, we turn to the autocorrelation of observables in a Markov Chain.  The autocorrelation of the estimator of an observable $\widehat{O}$ is defined as
\begin{equation}
    \Gamma_{\widehat{O}}(t) = \frac{1}{N} \sum_{t'=1}^N O(t') O(t'+t) - \langle \widehat{O} \rangle^2,
\end{equation}
where $O(t)$ is the estimator of the operator $\widehat{O}$ at time $t$ in the chain.  The integrated autocorrelation time for this observable is then defined as
\begin{equation}
    \tau_{\widehat{O}}^{\mathrm{int}} = 1 + 2\sum_{t=1}^{\infty} \frac{\Gamma_{\widehat{O}}(t)}{\Gamma_{\widehat{O}}(0)}.
    \label{eq:tau_int}
\end{equation}
As no Markov chain is employed in our direct sampling algorithm, the integrated autocorrelation time is equal to one.  For algorithms where a Markov chain does appear, we estimate the integrated autocorrelation time in Fig.~\ref{fig:ac}~(a).  We perform $10$ Markov chains where we measure the absolute magnetization after every update step.  For the direct sampling combined with Markov Chain and Wolff algorithms, we take a chain length of $N=10^4$ and for the MCMC algorithm we perform local updates and put a sample in the set after every sweep, i.e. $10^6$ local updates result in $N=10^4$ samples.  The autocorrelation function is averaged over the $10$ distinct chains.  We then calculate the integrated autocorrelation time by performing the sum up to a cut-off which we take as the time at which the autocorrelation function drops below a certain value, i.e. $\log(\Gamma_{\widehat{O}}(t))<-3$. Beyond that point, the correlation time becomes very noisy and the sum does not converge. The portion that is left out by doing this procedure can be estimated to be an exponentially decaying function.  Taking this exponentially decaying function in account leads to an underestimation of the integrated autocorrelation time by approximately $5\%$.  The MCMC algorithm has the largest autocorrelation time due to its local updates and relatively low acceptance ratio. As indicated in Fig.~\ref{fig:ac}~(b) our direct sampling + MC algorithm has higher acceptance ratio and global updates, the autocorrelation time is significantly lower.

\begin{figure}
\includegraphics{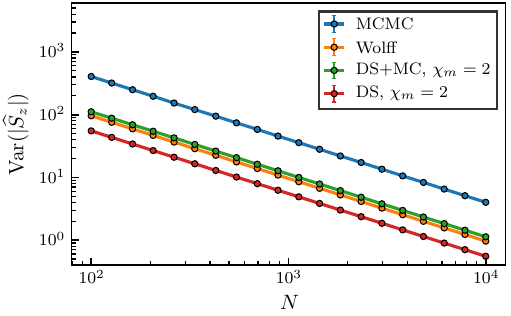}
\caption{Variance on the absolute magnetization of a $10 \times 10$ classical square-lattice Ising model as a function of sample set size, for different sampling algorithms.  The samples were drawn from the Boltzmann distribution at $kT/J = 1.9$.  All algorithms follow the expected $1/N$ behaviour.  MCMC has the largest variance, due to the large autocorrelation time.  The Wolff and direct sampling + MC algorithms have similiar variances.  The direct sampling algorithm without MC step has lowest variances.}
  \label{fig:variances}
\end{figure}
The autocorrelation time has an effect on the accuracy of estimates of observables, as it increases the variance of expectation values as determined by Eq.~\eqref{eq:var_psi}.  Specifically, it can be proven that the variance on an observable estimated using Markov Chain sampling is \cite{Gattringer2010}
\begin{align}
    \textrm{Var}_{p_{\Psi},\textrm{MCMC}}(\langle \widehat{O} \rangle) &= \frac{\tau_{\widehat{O}}^{\mathrm{int}}}{N} \textrm{Var}_{p_\Psi}(O_{loc}) \nonumber \\
    &= \frac{1}{N_{\mathrm{eff,MCMC}}} \textrm{Var}_{p_\Psi}(O_{loc}),
\end{align}
where we introduced the effective sample set size $N_{\mathrm{eff,MCMC}} = N/\tau_{\widehat{O}}^{\mathrm{int}}$.  The variance using a MCMC algorithm is thus $\tau_{\widehat{O}}^{\mathrm{int}}$ times higher than that obtained using a direct sampling algorithm of $p_{\Psi}$.  Conversely, in our direct sampling algorithm, we draw samples from a distribution $p_c$, and the importance sampling correction in the observable estimation of Eq.~\eqref{eq:qj_expectation} has a similar effect on the variance on that observable, defined in Eq.~\eqref{eq:var_pc}.  One can prove that the following relation exists \cite{Kong1992, Liu1996}:
\begin{align}
    \textrm{Var}_{p_{c}}(\langle \widehat{O} \rangle) &= \frac{\frac{1}{N} \sum_{\S \sim p_c} (|\Psi(\S)|^2/p_c(\S))^2}{N} \textrm{Var}_{p_\Psi}(O_{loc}) \nonumber \\
    &= \frac{1}{N_{\mathrm{eff,DS}}} \textrm{Var}_{p_\Psi}(O_{loc}),
    \label{eq:eff_DS}
\end{align}
where we introduced the effective sample set size $N_{\mathrm{eff,DS}} = N^2/\sum_{\S \sim p_c}(|\Psi(\S)|^2/p_c(\S))^2$.
From the above equation, it is clear that the variance of observables obtained using our direct sampling approach depends on the fluctuations in $|\Psi(\S)|^2/p_c(\S)$.   

Both MCMC and direct sampling have suboptimal variances due to the effective sample set sizes $N_{\mathrm{eff,MCMC}}$ and $N_{\mathrm{eff,DS}}$. To show their differences, we compare the variance on the absolute magnetization of a $10 \times 10$ classical Ising model for different number of samples $N$ using different sampling schemes in Fig.~\ref{fig:variances}.  All curves follow the expected scaling as $1/N$.  The vertical offset is determined by the effective sample set size, either the effect of $\tau_{\widehat{O}}^{\mathrm{int}}$ or the fluctuations in $|\Psi(\S)|^2/p_c(\S)$.  For the Markov-Chain based sampling schemes, the cluster updates in the Wolff algorithm and the accompanying low autocorrelation times yield the lowest variances.  Using our direct sampling scheme with $\chi_m=2$ yields little fluctuations in $|\Psi(\S)|^2/p_c(\S)$ and the effective sample set size is large, yielding even lower variances than the Wolff algorithm.  

\subsection{Effect of the marginal environment}

\begin{figure}
\includegraphics{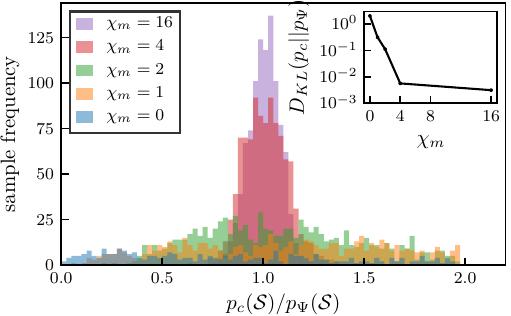}
\caption{Histograms of $p_c(\S)/p_{\Psi}(\S)$ for $1000$ samples drawn from $p_c$, for different cut-offs $\chi_m$ of the environments of the marginals.  The inset shows the Kullback-Leibler divergence between $p_c$ and $p_{\Psi}$.}
  \label{fig:marginal_environment}
\end{figure}

\begin{figure}
\includegraphics{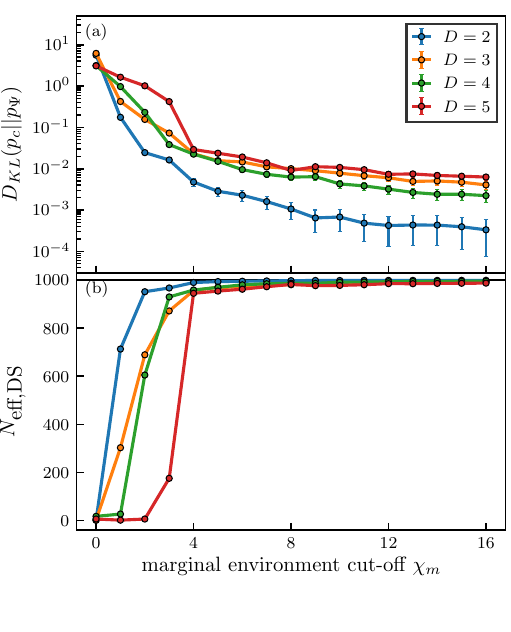}
\caption{Similarity between $p_c$ and $p_{\Psi}$. (a), Kullback-Leibler divergence $D_{KL}(p_c || p_{\Psi})$ as a function of the cut-off of the environments of the marginal probability distributions $\chi_m$ for PEPS wave functions optimized on the ground state of the AFH with different bond dimensions $D$.  The Kullback-Leibler divergence drops sharply for low values of $\chi_m$, signaling that the overlap between $p_c$ and $p_{\Psi}$ improves greatly with increasing $\chi_m$. (b), effective sample set size $N_{\textrm{eff,DS}}$ as defined in Eq.~\eqref{eq:eff_DS}, as a function of $\chi_m$.  The black line indicates the number of samples drawn, i.e. $N=1000$.  The effective sample set sizes reach values close to this optimum for $\chi_m \approx D$ and keep improving when increasing $\chi_m$.}
  \label{fig:DKL_Dscaling}
\end{figure}

As explained in Sec.~\ref{sec:ds}, the distribution $p_c$ approaches the distribution $p_{\Psi}$ systematically when increasing the cut-off $\chi_m$ of the environments of the marginal probability distributions.  In Fig.~\ref{fig:marginal_environment}, we show histograms of the ratio of the probabilities $p_c$ and $p_{\Psi}$ for different cut-offs $\chi_m$. We draw $1000$ samples using our direct sampling approach from a $10\times10$ PEPS with $D=4$, optimized on the ground state of the Antiferromagnetic Heisenberg Model (AFH) with Hamiltonian
\begin{equation}
    \wh{H} = \sum_{\langle i,j \rangle} \hat{S}_i^x \hat{S}_j^x + \hat{S}_i^y \hat{S}_j^y + \hat{S}_i^z \hat{S}_j^z,
    \label{eq:AFH}
\end{equation}
where $\hat{S}_i^{x,y,z}$ is the spin-$1/2$ operator on spin $i$ in the $x,y,z$-direction.
We use the notation $\chi_m=0$ to indicate that no environment was used, i.e. the unit MPO.  Direct sampling with $\chi_m=0$ leads to a large mismatch between $p_c$ and $p_{\Psi}$.  Using $\chi_m=1$, i.e. an environment that is still spatially uncorrelated but allowing non-trivial correlations between the virtual dimensions of the PEPS tensors, already improves the distance between both distributions.  For $\chi_m=4$, the histograms start to peak sharply around $p_c/p_{\Psi}=1$.  In the inset, we show the Kullback-Leibler divergence between $p_c$ and $p_{\Psi}$, defined as
\begin{equation}
    D_{\mathrm{KL}}(p_c || p_{\Psi}) = \sum_{\{\mathcal{S}\}} p_c(\mathcal{S}) \log \left( \frac{p_c(\mathcal{S})}{p_{\Psi}(\mathcal{S})} \right),
\end{equation}
which quantifies the relative entropy between the distributions $p_c$ and $p_{\Psi}$.  In Fig.~\ref{fig:DKL_Dscaling}~(a), we show $D_{\mathrm{KL}}(p_c || p_{\Psi})$ as a function of $\chi_m$ for PEPS wavefunctions with different bond dimensions $D$.  For values lower than $\chi_m \approx D$, the Kullback-Leibler divergence drops sharply, dropping up to three orders of magnitude compared to $\chi_m=0$.  In this region, the distribution $p_c/p_{\Psi}$ goes from a broad distribution to a narrow distribution centered around $p_c/p_{\Psi}=1$, as can be seen in Fig.~\ref{fig:marginal_environment}.  In Fig.~\ref{fig:DKL_Dscaling}~(b), we show the effective sample set size $N_{\textrm{eff,DS}}$ as a function of $\chi_m$.  When $\chi_m \approx D$, $N_{\textrm{eff,DS}}$ lies close to the number of samples drawn, indicating that variances obtained by sampling directly from $p_c$ are equally low as variances obtained by sampling from $p_{\Psi}$.
Note that, although a double layer contraction, the environments of the marginals can be calculated in advance and constitute a negligible cost compared to the cost of sampling for all shown $\chi_m$.  More generally, for the values of $\chi_m \approx D$ used here, the calculation of the marginals is $\mathcal{O}(D^2)$ more expensive than drawing a single sample.  For bond dimensions $D=1-10$ used in this work, this cost is lower than constructing all samples in the sample set, which is typically of size $N=1000-10000$.

\subsection{Boundary MPS accuracy}

\begin{figure}
\includegraphics{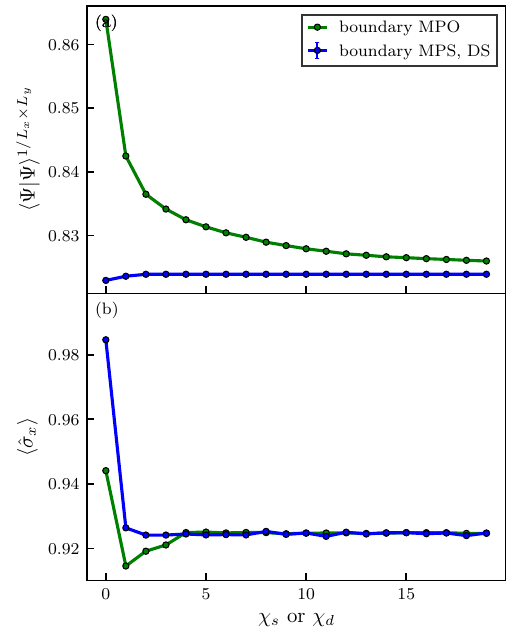}
\caption{Comparison between boundary MPS combined with direct sampling, and boundary MPO methods for calculating observables.  The observables are calculated for a $D=4$ ground state of a $11 \times 11$ Transverse Field Ising model at $g=3.2$.  For the sampled observables, $N=10^6$ samples were used to largely eliminate statistical fluctuations. (a), square norm per site of the PEPS.  The boundary MPS method converges more quickly than the boundary MPO method as a function of the cut-off.  (b), spin in transverse field direction of the center site.  The error in the environment of the boundary MPO method is largely canceled out as the environment appears in the numerator and the denominator of the expectation value.  Hence, the error on the expectation value converges more rapidly than the expectation value of the norm in panel (a).}
  \label{fig:obs_vs_chi}
\end{figure}
According to Eqs.~\eqref{eq:var_psi} and~\eqref{eq:var_pc}, sampling of observables introduces a variance on the resulting estimates.  Apart from this variance, the algorithm also introduces a bias, due to the cut-off $\chi_s$ of the single-layer boundary MPS.  This bias is also present in double-layer contractions, where it originates from the cut-off $\chi_d$ of the boundary MPO.  The bias can be systematically reduced by increasing $\chi_s$ or $\chi_d$.  The systematic reduction for both algorithms is shown in Fig.~\ref{fig:obs_vs_chi}.  In Fig.~\ref{fig:obs_vs_chi}~(a), we show the norm per site of the PEPS as a function of $\chi_s$ and $\chi_d$.  It is clear that the bias on the norm is small for the boundary MPS method, and converges rapidly to zero at small $\chi_s$.  The bias is markedly larger for the boundary MPO method, and converges more slowly to zero.  However, in typical expectation values containing local operators, the same environment enters the numerator and the denominator, hence largely canceling the errors on the environment.  This can be observed in Fig.~\ref{fig:obs_vs_chi}~(b), where the expectation value of the spin in the direction of the transverse field is shown as a function of $\chi_s$ or $\chi_d$.  The convergence of this expectation value for the boundary MPO algorithm is much faster than the convergence of the norm, although the norm is part of the calculation of this expectation value.  This effect is also present for the boundary MPS algorithm, but is to a lesser extent important as the unnormalized expectation values already converge for low $\chi_s$.  Comparing the convergence of the norm in Fig.~\ref{fig:obs_vs_chi}~(a) as a function of $\chi_d$, with the convergence of the effective sample set size $N_{\textrm{eff,DS}}$ in Fig.~\ref{fig:DKL_Dscaling}~(b) as a function of $\chi_m$, it is clear that $\chi_m$ can be taken much smaller than $\chi_d$, although they are both double layer contractions.  

All truncations were performed using singular value decompositions.  Based on these results and systematic investigations in Refs.~\cite{Liu2017GradientStates, Liu2019AccurateDimensions}, we use the heuristic $\chi_s=3D$ in all numerical experiments in this paper.  Fig.~\ref{fig:obs_vs_chi} shows that this is clearly sufficient.

\subsection{Optimizing ground states of quantum spin models}
In this section, we optimize a PEPS variationally using samples drawn according to our direct sampling algorithms. 

\subsubsection{Stochastic Reconfiguration versus Gradient Descent}
\label{sec:SRvsGD}
\begin{figure}
\includegraphics{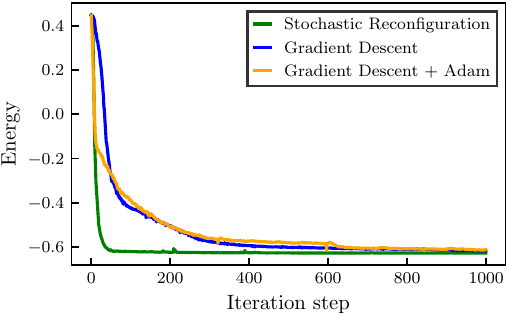}
\caption{Energy as a function of iteration steps for a $10\times10$ PEPS with $D=4$ using Stochastic Reconfiguration, Gradient Descent and Adam.  For all runs, expectation values were obtained with $N=1000$ samples.  All runs used a fixed step size, i.e. that for which convergence was fastest.  This is $0.05$ for SR and $0.01$ for SGD and Adam.  Note that SR requires an additional matrix inversion compared to SGD and Adam.  This step scales as $\mathcal{O}(D^4)$, i.e. more favourable than the cost of single-layer contractions.  In this case specifically, the matrix inversion adds roughly 30\% of runtime per iteration step.}
  \label{fig:SRvsGD}
\end{figure}
In Fig.~\ref{fig:SRvsGD}, we show how the energy decreases as a function of the number of optimization steps for the Stochastic Gradient Descent update (Eq.~\eqref{eq:gradient}), Adam update, and the Stochastic Reconfiguration update (Eq.~\eqref{eq:SR}), applied to finding the ground state of the antiferromagnetic Heisenberg model of Eq.~\eqref{eq:AFH}.
where $S_i^x$ is the spin operator in the $x$-direction applied on site $i$ and the sum runs over all pairs of nearest neighbors. All algorithms optimize the energy of a $D=4$ PEPS on a $10 \times 10$ square lattice, starting from a random initial PEPS with uniform positive parameters. For the stochastic reconfiguration, we use a diagonal shift $\delta=0.01$. The stochastic reconfiguration method is clearly advantageous to gradient descent or Adam, both in the number of iteration steps required to reach a certain accuracy and in the final accuracy the optimization can reach. 

\subsubsection{Energy convergence}

\begin{figure}
\includegraphics{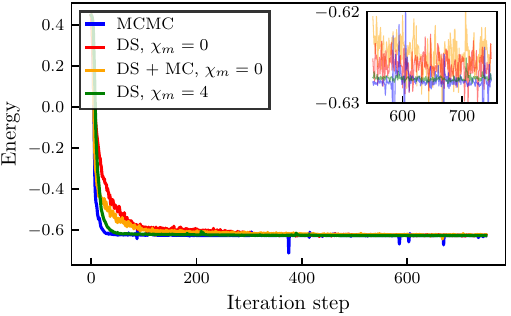}
\caption{Energy as a function of iteration steps for an optimization of the Heisenberg-model ground state using a $10 \times 10$, $D=4$ PEPS.  We compare the energy convergence for different sampling distributions, i.e. MCMC, direct sampling + Markov chain, and direct sampling with different $\chi_m$.  The inset shows a zoom of the last $200$ iteration steps, highlighting the variability of the energy between subsequent iteration steps and the final converged energies.  All runs use $N=1000$ samples to estimate expectation values and the MCMC algorithm uses $\tau_i=100$ equilibration sweeps and one sweep between successive samples.}
  \label{fig:energy_convergence}
\end{figure}

\begin{figure}
\includegraphics{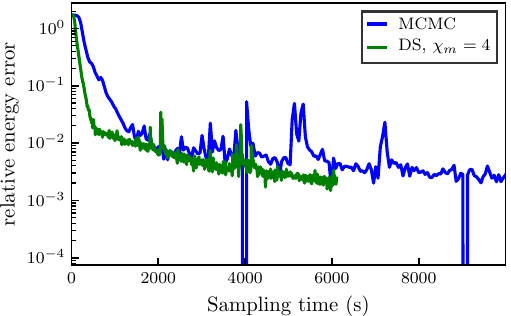}
\caption{Relative energy error as a function of wall time spent in the sampling routine for an optimization of the Heisenberg-model ground state using a $10 \times 10$, $D=4$ PEPS.  Both optimizations were performed with the same modest computational resources, i.e. parallelization over 8 CPU cores.  The MCMC algorithm consists of $\tau_i=100$ equilibration sweeps and one sweep between successive samples.}
  \label{fig:energy_convergence_time}
\end{figure}
We have shown the energy convergence as a function of iteration steps for Stochastic Gradient Descent, Adam and Stochastic Reconfiguration, and found that the latter converges much more rapidly.  A natural question to ask is whether the mismatch between $p_c$ and $p_{\Psi}$ also leads to differences in convergence.  In Fig.~\ref{fig:energy_convergence}, we show the energy convergence of a $10 \times 10$, $D=4$ PEPS using SR.  For the sampling, we use MCMC with $\tau_s=1$ and $\tau_i=100$ sweeps, direct sampling combined with a Markov chain with $\tau_s=1$ and $\tau_i=100$ samples, direct sampling with $\chi_m=0$ and direct sampling with $\chi_m=4$.  The first two yield samples according to $p_{\Psi}$ and the other two according to $p_c$.  From this figure, we see that sampling approaches with $\chi_m=0$ converge slower than MCMC and direct sampling with $\chi_m=4$.  Direct sampling with $\chi_m=4$, however, leads to the most stable convergence, and the energies reached are comparable with those obtained using MCMC. The convergence of MCMC suffers from unphysical peaks in energy, that can be traced back to poor mixing of the Markov-chain sampler. This problem is absent in the direct sampling approaches, as subsequent samples are uncorrelated from each other.  This leads to better statistics, also explaining the more stable convergence that can be noted in the inset of Fig.~\ref{fig:energy_convergence}.

In Fig.~\ref{fig:energy_convergence_time}, we show the relative energy error of the variational PEPS as a function of the time spent in the sampling routine.  We see that the convergence expressed in wall time of the direct sampling procedure is faster than that of MCMC.  Also in this figure, the effect of poor sampling sets in the MCMC procedure is visible.

\subsubsection{Comparison with other optimization algorithms}

\begin{figure}
\includegraphics{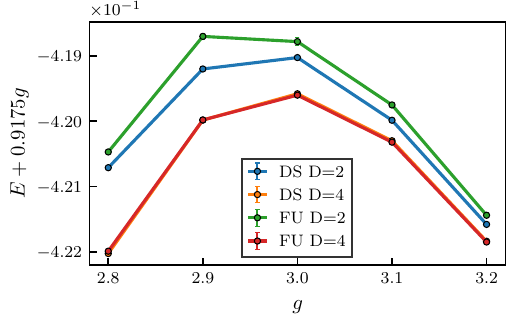}
\caption{Comparison of energies between direct sampling and full update for a $11 \times 11$ PEPS optimized on the TFI in the vicinity of the critical point.  We added a linear term in $g$ to the energy to make differences more visible.  For $D=2$, direct sampling performs better compared to FU, while both methods reach similar precision for $D=4$.  For the direct sampling results, $N=1000$ samples were used together with the SR update method.}
  \label{fig:ising_compare}
\end{figure}

To show the accuracy of direct sampling combined with gradient updates to optimize PEPS wavefunctions, we first compare our method with the full update (FU) method, which approximates imaginary-time evolution by Trotter-Suzuki gates and local truncations \cite{Lubasch2014AlgorithmsStates}. We consider a $11\times11$ PEPS with bond dimensions $D=2$ and $D=4$ and variationally find the ground state of the transverse field Ising model, with Hamiltonian
\begin{equation}
    \wh{H} = - \sum_{\langle i,j \rangle} \hat{S}_i^z \hat{S}_j^z - g \sum_{i} \hat{S}_i^x,
\end{equation}
where $g$ is a coupling constant quantifying the strength of the transverse field. The model has a critical point between a ferromagnetic ordered phase and a disordered phase in the vicinity of $g \approx 3$. In Fig.~\ref{fig:ising_compare}, we show the difference between energies obtained with the direct sampling approach and energies obtained with full update for different values of $g$. We see that for small bond dimension $D=2$, the direct sampling approach performs better than full update, especially in the vicinity of the critical point.

\begin{figure}
\includegraphics{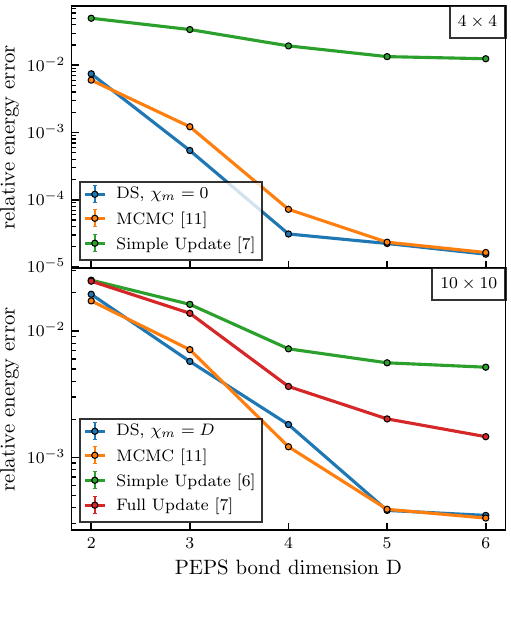}
\caption{Relative energy error $(E-E_{\mathrm{exact}})/E_{\mathrm{exact}}$ as a function of bond dimension on the ground state of the AFH model for different optimization algorithms and different lattice sizes.  An estimation of the exact energy is obtained through QMC. For the direct sampling results, $N=1000$ samples were used together with the SR update method.  Note that direct sampling and MCMC use the $U(1)$-projector to obtain samples with specific $U(1)$-charge, which may explain the difference in accuracy compared to simple and full update.}
  \label{fig:Dscaling}
\end{figure}
Second, we compare the accuracy of the ground state energy of the AFH obtained with direct sampling, with those obtained with simple update~\cite{Jiang2008}, full update~\cite{Lubasch2014AlgorithmsStates} and MCMC-based sampling~\cite{Liu2017GradientStates}.  In Fig.~\ref{fig:Dscaling}, we show how energies obtained with different optimization algorithms converge as a function of the PEPS bond dimension.  For the $4\times4$ system, we used $\chi_m=0$ in our sampling algorithm, and for the $10 \times 10$ system we used $\chi_m=D$.  Using our sampling approach, we obtain accuracies that are better than SU and FU.  Note however that the VMC results were obtained using the $U(1)$ symmetry, which is not the case for the SU and FU results.  The direct sampling results are consistent with MCMC sampling.  This shows that, although the sampling distribution $p_c$ is not equal to $p_{\Psi}$ and importance sampling needs to be used, the accuracy of the optimized wavefunctions does not suffer from the difference in sampling distribution.  By employing the environment of the marginal distributions, the distribution $p_c$ can be brought sufficiently close to $p_{\Psi}$ so that variances of sampled quantities are not affected.  This was already shown in Fig.~\ref{fig:marginal_environment}.  It has to be noted that in Ref.~\cite{Liu2017GradientStates}, it was reported that the sample set size required to obtain reliable gradients was $50 - 500$ times larger than the sample set sizes used in this work.  This may be explained by the uncorrelated nature of the samples drawn using direct sampling, compared to the existence of autocorrelation in Markov chain-based sampling approaches.  Another important difference with Ref.~\cite{Liu2017GradientStates} is that in our simulations the PEPS can be initialized randomly, compared with starting from a PEPS that is already optimized using the SU algorithm. The discrepancy in required sample set size may thus not be related to the direct sampling approach, but rather an indication of the importance of initialization in sampling-based tensor-network algorithms.

\section{Conclusion}
We have presented an algorithm for drawing uncorrelated samples from a probability distribution described by a double-layer PEPS tensor network.  This direct sampling algorithm was inspired by quantum jump Monte Carlo and can be performed in the single layer picture, thus scaling favorably with the bond dimension $D$ of the PEPS. The algorithm can be viewed as sampling a chain of conditional probability distributions, similar to the perfect sampling algorithm for MPS. The absence of a canonical form for a generic PEPS tensor network, however, implies that the marginal probability distributions cannot be constructed trivially. This forces us to employ importance sampling techniques as the sampled distribution $p_c$ is not equal to that defined by the square amplitudes of the PEPS, $p_{\Psi}$.  However, we provide a way to systematically reduce this discrepancy and show that the resulting algorithm provides meaningful samples.

Specifically, we showed by dimensionality reduction that the density of samples drawn according to our algorithm largely overlaps with the density of samples drawn using algorithms sampling $p_{\Psi}$.  We showed that using our samples as transitions in a Markov Chain results in small autocorrelation times and large acceptance rates.  Our sampling algorithm can be used in a VMC algorithm to determine ground states of quantum many-body systems.  The algorithm allows us to obtain expectation values efficiently and reliably, as is reflected in the accuracies we obtain on the Transverse Field Ising model and the Antiferromagentic Heisenberg model.  The resulting optimization is more stable than that using MCMC-based sampling when sample sets of comparable sizes are used.

We believe that our algorithm is useful for future studies using VMC in combination with PEPS wavefunctions. In particular, we expect that direct sampling will be more efficient than the use of MCMC in large-scale simulations, since the former can be parallellized indefinitely. We can extend our framework beyond ground-state simulations and consider excited states and time evolution; for the latter, we can use a stochastic implementation of the time-dependent variational principle.
\par One interesting open question concerns the sign structure of the variational PEPS wavefunction. In the case of neural-network inspired wavefunctions it has been observed that it is crucial to impose the correct sign structure on the wavefunction for the VMC optimization to be well-behaved \cite{Choo2019,westerhout2020generalization,szabo2020}. In our simulations of the nearest-neighbour Heisenberg model, we have observed that the optimization or variational result does not change when imposing a sign structure or not. It would be very interesting to apply our methods to frustrated systems (as in Ref.~\onlinecite{LiuGaplessModel}), for which the sign structure is not known a priori, and investigate this in detail.
\par Other applications include generative machine learning with PEPS tensor networks, which requires the ability to sample directly from a probability distribution defined by a single layer PEPS \cite{Han2018,Cheng2019}. 
\par A final interesting avenue is the combination of direct sampling with the class of isometric PEPS \cite{Zaletel2020IsometricDimensions}.  The particular structure of isometric PEPS renders the environment of the marginal distributions -- $E_m(\mathcal{S}_{>y})$ in Eq.~\eqref{eq:Em} --  trivial, and would allow us to directly sample from the distribution $p_{\Psi}$. This can be combined with VMC and variational optimization for unitary tensor networks manifolds~\cite{Hauru2020RiemannianNetworks} to find ground states that can be represented by isometric PEPS. We will investigate this avenue in future work.

Code to reproduce the results of this paper can be found at \url{https://github.com/tvieijra/DirectSamplingPEPS.jl}.
\begin{acknowledgements}
The authors would like to thank Sheng-Hsuan Lin and Frank Pollmann for inspiring discussions and valuable comments on our manuscript. 
This work has received funding from the European Research Council (ERC) under the European Unions Horizon 2020 research and innovation programme (grant agreements No 715861 (ERQUAF) and 647905 (QUTE)), and from Research Foundation Flanders (FWO) via grants GOE1520N and FWO18/ASP/279.
Computational resources (Stevin Supercomputer Infrastructure) and services used in this work were provided by the VSC (Flemish Supercomputer Center), and
the Flemish Government – department EWI.
\end{acknowledgements}

\bibliography{references}

\revappendix

\section{Quantum jump Monte Carlo}
\label{sec:app_QJMC}
Quantum jump Monte Carlo originated as a sampling approach for density matrices of systems connected to an environment subject to time evolution according to the Lindblad equation:
\begin{multline}
    \frac{d \rho}{dt} = -\frac{i}{\hbar}[H,\rho] \\ - \frac{1}{2} \sum_m \left( c_m^{\dag} c_m \rho + \rho c_m^{\dag} c_m - 2 c_m \rho c_m^{\dag} \right) .
\end{multline}
Here, $\rho$ is the density matrix of the system under study, $H$ is the Hamiltonian coupling the degrees of freedom within the system, and $c_m$ are Lindblad operators which describe the interaction of the system with the environment.  It is convenient to introduce the effective Hamiltonian $H_{\eff} = H - \frac{i}{2} \sum_m c_m^{\dag} c_m$, leading to the more compact equation
\begin{equation}
    \frac{d \rho}{dt} =  - i \left( H_{\eff} \rho -\rho H_{\eff}^{\dag} \right) + \sum_m c_m \rho c_m^{\dag}.
    \label{eq:master_app}
\end{equation}
At first order, the evolution of $\rho(t)$ can be written as
\begin{multline}
    \rho(t+dt) = \rho(t)  - i dt (H_{\eff} \rho -\rho H_{\eff}^{\dag}) \\ + dt \sum_m c_m \rho c_m^{\dag}.
    \label{eq:ev}
\end{multline}
The density operator $\rho(t)$ can be evolved stochastically using the following algorithm.
\begin{enumerate}
    \item At time $t=0$, sample a pure state $\ket{\phi(t=0)}$ from the mixed density matrix $\rho(t=0) = \sum_{\{\phi\}}p_{\phi} \rho_{\phi}(0)$, where $\rho_{\phi}(0) = \ket{\phi(0)}\bra{\phi(0)}$ and the sum runs over all pure states in the mixture.
    \item For a small increment $dt$, evolve $\ket{\phi(t)}$ using one of the following two ways, where $dp = \sum_m\braket{\phi(t) | c_m^{\dag} c_m | \phi(t)} = \sum_m{dp_m}$:
    \begin{itemize}
        \item With probability $1-dp$, evolve using 
        \begin{equation}
            \ket{\phi(t+dt)} = \frac{1 - iH_{\eff}dt}{\sqrt{1-dp}} \ket{\phi(t)}.
        \end{equation}
        \item With probability $dp$, evolve using 
        \begin{equation}
            \ket{\phi(t+dt)} = \frac{c_m}{\sqrt{dp_m/dt}} \ket{\phi(t)},
        \end{equation}
        where a particular $c_m$ is chosen according to the probability distribution $dp_m/dp$.
    \end{itemize}
    \item Repeat step two until the desired time $t$ is reached.
\end{enumerate}
On the level of the density operator $\rho_{\phi}(t+dt)$, resulting from evolving the pure state $\ket{\phi(t)}$, the evolution procedure leads to
\begin{align}
    & \rho_{\phi}(t+dt) = \ket{\phi(t)}\bra{\phi(t)} \nonumber \\
    & \qquad - i dt \left( H_{\eff} \ket{\phi(t)}\bra{\phi(t)}  - \ket{\phi(t)}\bra{\phi(t)}H_{\eff}^{\dag} \right) \nonumber \\
    & \qquad + dt \sum_m c_m \ket{\phi(t)}\bra{\phi(t)} c_m^{\dag}.
\end{align}
The sampling over the initial mixture of pure states leads then to the properly evolved density matrix $\rho(t) = \sum_{\{\phi\}}p_{\phi} \rho_{\phi}(t)$.

\section{Sampling of density matrix}
\label{density_sampling}

In the sampling algorithm outlined in the main text, we need to sample from the distribution
\begin{equation}
p_c(\mathcal{S}_y|\mathcal{S}_{<y}) = \braket{\phi_{y-1}|_b c(\mathcal{S}_y)^{\dag} c(\mathcal{S}_y)|\phi_{y-1}}_b,
\end{equation}
in the case that $\chi_m=0$, or
\begin{equation}
p_c(\mathcal{S}_y|\mathcal{S}_{<y}) = \braket{\phi_{y-1}|_b c(\mathcal{S}_y)^{\dag} E_m(\mathcal{S}_{>y}) c(\mathcal{S}_y)|\phi_{y-1}}_b ,
\end{equation}
in the case that the environment of the marginal is used. For notational simplicity, we will use the latter notation, assuming the MPO $E_m(\mathcal{S}_{>y})$ is the unit MPO when $\chi_m=0$ is used. In diagrammatic notation, this probability can be drawn as
\begin{align}
p_c(\mathcal{S}_y|\mathcal{S}_{<y}) = 
\begin{array}{c}
    \begin{tikzpicture}[scale=.5]
    \foreach \x in {0,2,...,6}{
    \draw[color=white,myBG,shift={(\x,0)}] (0,0) -- (0,-4);
    \draw[shift={(\x,0)}] (0,0)--(0,-4);
    };
    \draw[color=white,myBG,shift={(0,0)}] (0,0) -- (6,0);
    \draw[shift={(0,0)},line width=1] (0,0) -- (6,0);
    \draw[color=white,myBG,shift={(0,-2)}] (0,0) -- (6,0);
    \draw[shift={(0,-2)},line width=1] (0,0) -- (6,0);
    \draw[color=white,myBG,shift={(0,-4)}] (0,0) -- (6,0);
    \draw[shift={(0,-4)},line width=1] (0,0) -- (6,0);
    \foreach \x in {0,2,...,6}{
        \draw[shift={(\x,0)}]  (0,0)--(1,1);
        \draw[shift={(\x,-4)}]  (0,0)--(1,-1);
        \filldraw[tenwhite,shift={(\x+1,+1)}] (0,0) circle (0.40);
        \filldraw[tenwhite,shift={(\x+1,-5)}] (0,0) circle (0.40);
        \foreach \y in {-4,0}{
    	    \filldraw[tenorange,shift={(\x,\y)}]  (-1/2,-1/2)--(1/2,-1/2)--(1/2,1/2)--(-1/2,1/2)--(-1/2,-1/2);
	    };
	    \filldraw[tengreen,shift={(\x,-2)}]  (-1/2,-1/2)--(1/2,-1/2)--(1/2,1/2)--(-1/2,1/2)--(-1/2,-1/2);
    };
    \node at(0+1,1) {$s_1$};    
    \node at(2+1,1) {$s_2$};
    \node at(4+1,1) {$s_3$};
    \node at(6+1,1) {$s_4$};
    \node at(0+1,-5) {$s_1$};    
    \node at(2+1,-5) {$s_2$};
    \node at(4+1,-5) {$s_3$};
    \node at(6+1,-5) {$s_4$};
    \end{tikzpicture}
    \end{array}
    \label{eq:p_samp},
\end{align}
where we define $\mathcal{S}_y \equiv (s_1,...,s_N)$ for notational simplicity.
\par A multidimensional probability distribution $p(s_1,s_2,s_3,s_4)$ such as the one in Eq.~\eqref{eq:p_samp} can always be decomposed in a product of conditional probabilities using the bayesian rule:
\begin{multline} \label{eq:autoregressive}
    p(s_1,s_2,s_3,s_4) \\ = p(s_4|s_3,s_2,s_1)p(s_3|s_2,s_1)p(s_2|s_1)p(s_1),
\end{multline}
where the marginal probability distribution is given by
\begin{equation}
    p(s_1) = \sum_{s_1,s_2,s_3}p(s_1,s_2,s_3,s_4),
\end{equation}
and $p(x|y)$ is the conditional probability distribution of the set of degrees of freedom $x$, conditioned on the set of degrees of freedom $y$. Sampling from $p(s_1,s_2,s_3,s_4)$ can now be done sequentially, by first sampling a configuration $(s_1)$ from $p(s_1)$, then using the outcome to construct $p(s_2|s_1)$ and sample from this the configuration $(s_1,s_2)$, and repeating this until a configuration of all degrees of freedom is sampled.
\par Any probability distribution can be written in this form, but calculating the marginal probability distributions is in principle exponentially hard. To sample this probability distribution, we adapt the perfect sampling approach for unitary tensor networks \cite{Ferris2012PerfectNetworks}. To this end, we first calculate $p(s_1)$ defined in Eq.~\eqref{eq:autoregressive} as the marginal over all trailing degrees of freedom except the first: 
\begin{align}
p_{c}(s_1|\mathcal{S}_{<y}) &= \sum_{\{s_{2} ... s_{4}\}}  p_c(\mathcal{S}_{y}|\mathcal{S}_{<y}) \nonumber \\
&= 
\begin{array}{c}
    \begin{tikzpicture}[scale=.5]
    \draw[color=white,myBG,shift={(0,0)}] (0,0) -- (0,-4);
    \draw[shift={(0,0)}] (0,0)--(0,-4);
    \draw[shift={(0,-2)}] (0,0) -- (2,0);
    \draw[shift={(0,0)}]  (0,0)--(1,1);
    \draw[shift={(0,-4)}]  (0,0)--(1,-1);
    \filldraw[tenwhite,shift={(0+1,+1)}] (0,0) circle (0.40);
    \filldraw[tenwhite,shift={(0+1,-5)}] (0,0) circle (0.40);
    \foreach \y in {-4,0}{
	    \filldraw[tenorange,shift={(0,\y)}]  (-1/2,-1/2)--(1/2,-1/2)--(1/2,1/2)--(-1/2,1/2)--(-1/2,-1/2);
    };
    \filldraw[tengreen,shift={(0,-2)}]  (-1/2,-1/2)--(1/2,-1/2)--(1/2,1/2)--(-1/2,1/2)--(-1/2,-1/2);
    \draw (1/2,0) to[out=0, in=0, distance=2cm] (1/2,-4);
    \filldraw[tengrey,shift={(2,-2)}] (0,0) circle (0.4);
	\node at(2,-2) {$e_1$};
    \node at(0+1,1) {$s_1$}; 
    \node at(0+1,-5) {$s_1$}; 
    \end{tikzpicture}
\end{array}.
\label{eq:right_orth}
\end{align}

Here, we defined the environment $e_1$ obtained by contracting all columns of the tensor network defined in Eq.~\eqref{eq:p_samp}, where we also contract over the physical degrees of freedom to perform the marginalization.  The contraction in Eq.~\eqref{eq:right_orth} yields (the diagonal of) a density matrix, from which we sample $s_1$ according to the probabilities defined as its entries.  As this is a low-dimensional distribution, we can make use of simple algorithms yielding uncorrelated samples.  Once we have sampled $s_1$, we can construct the probability distribution $p(s_2|s_1)$ as follows
\begin{align}
p_{c}(s_2|\mathcal{S}_{\leq1,<y}) &= \sum_{s_3,s_4} p_c(s_2,s_3,s_4|\mathcal{S}_{\leq 1,<y}) \nonumber \\
&=
\begin{array}{c}
    \begin{tikzpicture}[scale=.5]
    \draw[color=white,myBG,shift={(0,0)}] (0,0) -- (0,-4);
    \draw[shift={(0,0)}] (0,0)--(0,-4);
    \draw[shift={(0,-2)}] (0,0) -- (2,0);
    \draw[color=white,myBG,shift={(2,0)}] (0,0) -- (0,-4);
    \draw[shift={(2,0)}] (0,0)--(0,-4);
    \draw[shift={(2,-2)}] (0,0) -- (2,0);
    \draw[color=white,myBG,shift={(0,0)}] (0,0) -- (2,0);
    \draw[shift={(0,0)}] (0,0) -- (2,0);
    \draw[color=white,myBG,shift={(0,-2)}] (0,0) -- (2,0);
    \draw[shift={(0,-2)}] (0,0) -- (2,0);
    \draw[color=white,myBG,shift={(0,-4)}] (0,0) -- (2,0);
    \draw[shift={(0,-4)}] (0,0) -- (2,0);
    \draw[shift={(0,0)}]  (0,0)--(1,1);
    \draw[shift={(0,-4)}]  (0,0)--(1,-1);
    \filldraw[tenwhite,shift={(0+1,+1)}] (0,0) circle (0.40);
    \filldraw[tenwhite,shift={(0+1,-5)}] (0,0) circle (0.40);
    \draw[shift={(2,0)}]  (0,0)--(1,1);
    \draw[shift={(2,-4)}]  (0,0)--(1,-1);
    \filldraw[tenwhite,shift={(2+1,+1)}] (0,0) circle (0.40);
    \filldraw[tenwhite,shift={(2+1,-5)}] (0,0) circle (0.40);
    \foreach \y in {-4,0}{
	    \filldraw[tenorange,shift={(0,\y)}]  (-1/2,-1/2)--(1/2,-1/2)--(1/2,1/2)--(-1/2,1/2)--(-1/2,-1/2);
    };
    \foreach \y in {-4,0}{
	    \filldraw[tenorange,shift={(2,\y)}]  (-1/2,-1/2)--(1/2,-1/2)--(1/2,1/2)--(-1/2,1/2)--(-1/2,-1/2);
    };
    \filldraw[tengreen,shift={(0,-2)}]  (-1/2,-1/2)--(1/2,-1/2)--(1/2,1/2)--(-1/2,1/2)--(-1/2,-1/2);
    \filldraw[tengreen,shift={(2,-2)}]  (-1/2,-1/2)--(1/2,-1/2)--(1/2,1/2)--(-1/2,1/2)--(-1/2,-1/2);
    \draw (2+1/2,0) to[out=0, in=0, distance=2cm] (2+1/2,-4);
    \filldraw[tengrey,shift={(4,-2)}] (0,0) circle (0.4);
	\node at(4,-2) {$e_2$};
    \node at(0+1,1) {$s_1$}; 
    \node at(0+1,-5) {$s_1$}; 
    \node at(2+1,1) {$s_2$}; 
    \node at(2+1,-5) {$s_2$};
    \end{tikzpicture}
\end{array} \nonumber \\
&= 
\begin{array}{c}
    \begin{tikzpicture}[scale=.5]
    \draw[color=white,myBG,shift={(2,0)}] (0,0) -- (0,-4);
    \draw[shift={(2,0)}] (0,0)--(0,-4);
    \draw[shift={(2,-2)}] (-2,0) -- (2,0);
    \draw[shift={(2,0)}]  (0,0)--(1,1);
    \draw[shift={(2,-4)}]  (0,0)--(1,-1);
    \filldraw[tenwhite,shift={(2+1,+1)}] (0,0) circle (0.40);
    \filldraw[tenwhite,shift={(2+1,-5)}] (0,0) circle (0.40);
    \foreach \y in {-4,0}{
	    \filldraw[tenorange,shift={(2,\y)}]  (-1/2,-1/2)--(1/2,-1/2)--(1/2,1/2)--(-1/2,1/2)--(-1/2,-1/2);
    };
    \filldraw[tengreen,shift={(2,-2)}]  (-1/2,-1/2)--(1/2,-1/2)--(1/2,1/2)--(-1/2,1/2)--(-1/2,-1/2);
    \draw (2+1/2,0) to[out=0, in=0, distance=2cm] (2+1/2,-4);
    \filldraw[tengrey,shift={(4,-2)}] (0,0) circle (0.4);
	\node at(4,-2) {$e_2$};
	\draw (1+1/2,0) to[out=180, in=180, distance=2cm] (1+1/2,-4);
    \filldraw[tengrey,shift={(0,-2)}] (0,0) circle (0.4);
	\node at(0,-2) {$\sigma_1$};
    \node at(2+1,1) {$s_2$}; 
    \node at(2+1,-5) {$s_2$};
    \end{tikzpicture}.
\end{array}
\end{align}
Here, we require the environment $e_2$, which is defined in a similar manner as $e_1$, but now contracting only over all but the first two columns in Eq.~\eqref{eq:p_samp}. Note that during calculation of $e_1$, we already calculated all other environments $e_{2...4}$. Furthermore, we compute the environment $\sigma_1$ by contracting the first column of Eq.~\eqref{eq:p_samp}, where the orange tensors are projected on the value of $s_1$ sampled in the previous step.
The resulting contraction yields a density matrix from which we can sample $s_2$ using the probabilities specified on the diagonal.  We can keep repeating this procedure until we have sampled all the degrees of freedom.  In total, this yields an uncorrelated configuration from the probability distribution $p_c(\mathcal{S}_y)$.

\end{document}